\documentclass{iopart}
\usepackage{iopams}
\usepackage{amssymb}
\usepackage{graphicx}
\graphicspath{ {FigFiles/} } %% tidy subdir for figure files
\usepackage{bbold}   %% for a nice identity symbol 
\usepackage{natbib}		%% to make references easier
\setcitestyle{authoryear}	%% use {square,numbers} for numerical citations
\usepackage[breaklinks=true,%
            colorlinks=true,%
            linkcolor=blue,%
            urlcolor=blue,%
            citecolor=blue]{hyperref}
\usepackage{enumitem}
\usepackage{subfigure}

\newcommand{\bra}[1]{\ensuremath{\left\langle#1\right|}}
\newcommand{\ket}[1]{\ensuremath{\left|#1\right\rangle}}
\newcommand{\braket}[2]{\ensuremath{\left\langle#1\middle\vert#2\right\rangle}}
\newcommand{\ketbra}[2]{\ensuremath{\left|#2\right\rangle\left\langle#1\right|}}
\newcommand{\binom}[2]{{\ensuremath{{#1}\choose{#2}}}} % seems not in iopams
\usepackage{datetime} % for time stamping current version
\usepackage[textsize=footnotesize,backgroundcolor=red!25]{todonotes}

\definecolor{armygreen}{rgb}{0.29, 0.33, 0.13}

\begin{document}
\title[Quantum walk spin glass ground states]{Finding spin glass ground states using quantum walks}

\author{Adam Callison$^1$, Nicholas Chancellor$^2$, Florian Mintert$^1$ and Viv Kendon$^2$}
\address{$^1$ Blackett Laboratory, Imperial College London, London SW7~2BW, UK}
\address{$^2$ Physics Department, Durham University, South Road, Durham, DH1~3LE, UK}
\eads{\mailto{viv.kendon@durham.ac.uk}, \mailto{a.callison16@ic.ac.uk}}
%%%%%%%%%%%%%%%%
\begin{abstract}
Quantum computation using continuous-time evolution under a natural hardware Hamiltonian is a promising near- and mid-term direction toward powerful quantum computing hardware. 
We investigate the performance of continuous-time quantum walks as a tool for finding spin glass ground states, a problem that serves as a useful model for realistic optimization problems.
By performing detailed numerics, we uncover significant ways in which solving spin glass problems differs from applying quantum walks to the search problem. 
Importantly, unlike for the search problem, parameters such as the hopping rate of the quantum walk do not need to be set precisely for the spin glass ground state problem. 
Heuristic values of the hopping rate determined from the energy scales in the problem Hamiltonian are sufficient for obtaining a better quantum advantage than for search.
We uncover two general mechanisms that provide the quantum advantage: matching the driver Hamiltonian to the encoding in the problem Hamiltonian, and an energy redistribution principle that ensures a quantum walk will find a lower energy state in a short timescale.
This makes it practical to use quantum walks for solving hard problems, and opens the door for a range of applications on suitable quantum hardware.
\end{abstract}
%%%%%%%%%%%%%%
%\noindent{\textcolor{red}{\small \textit{Draft compiled at \currenttime \,\, \today}}}


%\submitto{\NJP}
%%%%%%%%%%
\maketitle
%%%%%%%%%%%%%%%%%%%%%%%%%%%%% main text starts here %%%%%%%%%%%%%%%%%%%%%%%%%%%
\tableofcontents	% useful while drafting if not in final version

%------------------------------------------------------------------------%
\section{Introduction}\label{sec:intro}
%------------------------------------------------------------------------%
\markboth{\textit{\rightmark}}{\thepage}

Optimization problems need to be solved in a broad range of areas, such as scheduling, route planning, supply chains, finance.  
This is often computationally intensive, so the prospect of quantum enhanced solution methods is an important research direction for practical quantum computing.  
One way to tackle optimization in a quantum setting is to use a device which realises an Ising Hamiltonian with a transverse field.
Computing using the Ising Hamiltonian works as follows:
The optimization problem is encoded into the Ising Hamiltonian $\hat{H}_I$ 
\begin{eqnarray}\label{eq:Ising_long}
    \hat{H}_I = -\sum_{(j\neq k)=0}^{n-1}J_{jk}\hat{Z}_j\hat{Z}_k - \sum_{j=0}^{n-1}h_j\hat{Z}_j,
\end{eqnarray}
on $n$ qubits, such that the solution corresponds to the ground state of $\hat{H}_I$.
In our notation, the operator $\hat{Z}_j$ on the full Hilbert space applies the single qubit Pauli-$Z$ operator $\hat{Z}$ to the $j$th qubit,
\begin{equation}\label{eq:fullZdef}
    \hat{Z}_j =  
    \left(\bigotimes_{r=0}^{j-1} \hat{\mathbb{1}}_2\right) 
    \otimes \hat{Z} \otimes
    \left(\bigotimes_{r=j+1}^{n-1} \hat{\mathbb{1}}_2\right),
\end{equation}
where $\hat{\mathbb{1}}_2$ is the identity operator on a single qubit.
The (real) values of the coupling strengths $J_{jk}$ and fields $h_j$ define the optimization problem, and efficient methods are known for expressing optimization problems in terms of these coupling and field strengths \citep[e.g.,][]{choi2010adiabatic}.
The transverse field term $\hat{H}_T$ 
\begin{eqnarray}\label{eq:Ising_trans}
    \hat{H}_T = -\Gamma\sum_{j=0}^{n-1}\hat{X}_j,
\end{eqnarray}
drives transitions between states,
where $\Gamma$ is a real-valued transverse field strength, and $\hat{X}_j$ is the operator on the full Hilbert space that applies the single qubit Pauli-$X$ operator to the $j$th qubit,  defined by analogy with $\hat{Z}_j$ in (\ref{eq:fullZdef}).
The qubits are initialised in the ground state of $\hat{H}_T$, this is easy to do by applying a strong transverse field to align all the qubits in the state $\ket{+} = 2^{-1/2}(\ket{0}+\ket{1})$.
Then, the computation is carried out by applying the full transverse Ising Hamiltonian
\begin{eqnarray}\label{eq:TIM}
    \hat{H}_\mathrm{TI}(t) = A(t)\hat{H}_T + B(t)\hat{H}_I,
\end{eqnarray}
where $t$ is time and $A(t)$, $B(t)$ are real-valued control functions. 
To obtain a candidate solution to the optimization problem, the qubit register is measured after a time $t_f$.  
For some problems, sampling from the distribution of low energy states provides the required solution -- this can be done by repeating the computation, which will in general not produce the lowest energy state with certainty.

The Ising Hamiltonian is a natural choice for encoding problems for two reasons. 
First, it is proven to be universal for classical problems \citep{de2016simple}.
There are efficient methods for mapping NP-hard optimization problems to the Ising model \citep{lucas2014ising,choi2010adiabatic}, providing a practical route to quantum algorithms.  
Since many optimization problems are NP-hard, an exponential speed up is not expected, but even modest polynomial improvements are useful for practical applications. 
There is increasing interest in how to obtain polynomial advantages through quantum algorithms \citep{Moylett2017tsp,Montanaro2015,ambainis2019}.
Interesting results have been presented for a wide range of applications, such as mathematics \citep{Bian13a,Li17a}, computer science \citep{chancellor16a}, computational biology \citep{perdomo-ortiz12a}, finance \citep{marzec16a}, and aerospace \citep{coxson14a}.
Second, the Ising Hamiltonian can be implemented in a range of different physical systems. 
The quantum Ising Hamiltonian is the basic interaction Hamiltonian in the D-Wave Systems Inc.~programmable superconducting devices \citep{D-wave,boixo2013experimental,johnson2011quantum}.
Implementations in other promising architectures include Rydberg systems \citep{Bernien17a} and trapped ions \citep{Kim2011ionIsing}.  
The Ising Hamiltonian is also the basic tool for specialised optimization hardware, such as coherent Ising machines \citep{Inagaki16a,McMahon16a}. 
Optimization using the Ising Hamiltonian can be implemented in digital quantum architectures by using the quantum approximate optimization algorithm (QAOA) \citep{Farhi14a,Farhi14b,Marsh2018qwqaoa} or quantum alternating operator ansatz \citep{Hadfield17a}.  Studies by \cite{zhou2018qaoa} show how to exploit non-adiabatic effects in QAOA on early quantum hardware.

There are several known methods for driving the quantum system from its initial state into the ground state of a Hamiltonian defining the problem to be solved.  
These methods correspond to different choices for the control functions $A(t)$ and $B(t)$ in (\ref{eq:TIM}). 
Adiabatic quantum computing \citep{Kado1998,farhi2000quantum,farhi2001quantum} keeps the quantum system in the ground state while the initial Hamiltonian is slowly changed into the problem Hamiltonian.
Quantum annealing \citep{Fini1994} takes advantage of open quantum systems effects to cool the system towards the ground state. 
Continuous-time quantum walks evolve the system under a time-independent Hamiltonian for a suitable time before measurement of the final state. 
Computation by continuous-time quantum walk and adiabatic quantum computing are end points of a family of continuous-time protocols that use the same Hamiltonian terms but are applied with different time dependent modulation \citep{morley2017quantum}.
In this work, we focus on computation by quantum walk using time-independent transverse Ising Hamiltonians.

Quantum walks can solve the search problem \citep{childs2004spatial}, achieving the same quadratic $O(N^{1/2})$ quantum speed up as is obtained by Grover's algorithm \citep{grover1996fast}. We describe the search problem further in Subsection \ref{ssec:qsearch}.
For particular graphs, quantum walks can solve problems exponentially faster \citep[e.g.,][]{childs2003exponential}, and quantum walks are now widely used as subroutines in more complex quantum algorithms.  
However, in the continuous-time setting, the application of quantum walks to optimization problems has not been studied in detail.
There is increasing interest in quenches \citep{Amin18a} or pauses \citep{Marshall19a,passarelli19a} in quantum annealing, which effectively run an open-system version of a quantum walk during part of the computation. 
Thermal relaxation effects dominate in the regime currently accessible by flux qubit quantum annealers, which is the focus of these works. 
An algorithm which is essentially a quantum walk on a spin glass, although presented using different terminology, has been analysed by \cite{Hastings19a}.
Along with the same energy conservation arguments we describe in section \ref{ssec:energycons}, Hastings' findings suggest that quantum walks on spin glasses will be interesting to explore.
Given that quantum walks provide a better performance for searching than adiabatic quantum computing, especially when limited coherence time and other practical factors, such as precision of control settings, are considered \citep{morley2017quantum}, it is important to understand how they perform for a wider range of problems.

In this work, we tackle the question of if, and how, a quantum walk can be useful for practical quantum optimization.  
We present a detailed numerical investigation of continuous-time quantum walks applied to solving combinatorial optimization problems, using the Sherrington-Kirkpatrick spin glass ground state problem as a prototypical example.
Finding the ground state of a frustrated Sherrington-Kirkpatrick spin glass \citep{kirkpatrick1975solvable} is known to be not only NP-hard, but also \emph{uniformly}-hard, as suggested by its finite-temperature spin glass transition. Without a finite temperature spin glass transition, a problem cannot be uniformly hard, since the lack of a transition implies that typical cases will be easy for the Monte Carlo family of algorithms, as discussed in \citep{katzgraber2014glassy}.
As has been shown for a random problem type used in early benchmarks of quantum annealing hardware \citep{katzgraber2014glassy}, uniform hardness is crucial: without this property, \emph{randomly} generated instances of NP-hard problems are not necessarily hard to solve \citep{Beier2004a,Krivelevich2006a,lucas2014ising}. 

We use a random energy model \citep{derrida1980random} for comparisons, to draw out the effects of the correlations between energy difference and Hamming distance in the spin glass.  
A problem with perfect correlations is easy to solve, like finding the ground state of a spin system with only local fields, no couplings.
A completely random problem, such as finding the ground state of a random energy model instance, has no correlation to exploit and so is very hard to solve, essentially requiring random guessing. 
However, a completely random model is fully characterised by average values of its properties, and finding exact ground states of specific instances is typically not interesting.
Intermediate problems with some correlations are both hard and interesting, with complex behaviour and phase diagrams, like spin models with frustration and spin glass phases.
Real optimization problems typically have correlations; they are often hard to solve but also produce interesting solutions.  The inherent complexity of a problem comes from the structures of the problem and its correlations, not the structure of the solution itself.  
One illustration of this is the construction of hard benchmarking problems with `planted' solutions defined at the time of construction, which therefore have no special structure related to the problem's hardness, see for example \citep{Hen19planted,Hamze19tunable}.

The paper is structured as follows: 
In section \ref{sec:QWcomp}, we review the setting for computation by continuous-time quantum walk encoded into qubits, including application to the search problem. 
In section \ref{sec:spinglassproblems}, we introduce the Sherrington-Kirkpatrick spin glass model, and the random energy model we use for comparison. 
In section \ref{sec:numericalmethods}, we describe the numerical methods used in this investigation. 
In section \ref{sec:qw_on_sg}, we present the main results showing how quantum walks can find spin glass ground states more effectively than a quantum search algorithm. 
In section \ref{sec:compmech}, we identify the computational mechanisms and important aspects of the problem structure that contribute to the effectiveness of quantum walk computation. 
Finally, in section \ref{sec:outlook}, we summarize and conclude.

%%%%%%%%%%%%%%%%%%%%%%%%%%%%%%%%%%%%%%%%%%%%%%%%%%%%%%%%%%%%%%%%%%%%%%%%%%
\section{Computing with quantum walks}\label{sec:QWcomp}
%------------------------------------------------------------------------%

Both discrete (coined) quantum walks  \citep{aharonov00a,shenvi2003quantum} and continuous-time quantum walks \citep{farhi98a,childs2003exponential} are used for computation.   
This work only uses the continuous-time quantum walk, and also only as an \emph{encoded} quantum walk, in which qubits are used to store the binary labels of the positions of the quantum walker (see figure \ref{fig:hypercube} for a simple example).   

%------------------------------------------------------------------------%
\subsection{Continuous-time quantum walks}\label{ssec:qwcont}
%------------------------------------------------------------------------%

A continuous-time quantum walk is defined on an undirected graph $G(V,E)$, with $V=\{j\}_{j=0}^{N-1}$ the set of $N$ vertex labels and $E$ the set of label-pairs $(j,k)$ associated with edges. 
The vertices correspond to the positions of the walker, and the edges indicate the allowed transitions between vertices.
This is conveniently encoded in the adjacency matrix $A$ of the graph, which has entries $A_{jk} = 1$ for $(j,k) \in E$ and $A_{jk} = 0$ otherwise. 
The Laplacian of $G$ is $L=A-D$, where $D$ is a diagonal matrix formed from the degree of each vertex, $D_{jj}=\mbox{deg}(j)$, where $\mbox{deg}(j)$ is the number of edges connected to vertex $j$. 
Both the adjacency matrix $A$ and Laplacian $L$ are symmetric matrices which can thus be used to define a quantum Hamiltonian for the dynamics of the continuous-time quantum walk on the graph. 
In this work, we only need regular graphs, for which $\mbox{deg}(j)$ is constant with respect to $j$. 
For regular graphs, the only difference between using the adjacency matrix $A$ or Laplacian $L$ is an irrelevant global phase \citep{childs2004spatial}. 
We use the Laplacian form of the Hamiltonian for consistency with prior work. 
We thus define the quantum walk Hamiltonian $\hat{H}_{G}$ for a quantum walk on graph $G$ by
\begin{equation}\label{eq:qwcont}
    \bra{j}\hat{H}_{G}\ket{k} = -\gamma L_{jk},
\end{equation}
where $\gamma$ is the hopping rate between connected vertices per unit time. 
The states $\ket{j},\ket{k}$ for $j,k\in V$ are associated with the vertices of $G$ and form a basis for a Hilbert space of dimension $N$.
In the Ising model context, the dimension of the Hilbert space is $N=2^n$ where $n$ is the number of qubits, and $\{|j\rangle\}_{j=0}^{N-1}$ is the computational basis.
For a quantum walk starting in state $\ket{\psi(0)}$, the state of the walker evolves according to the Schr{\"o}dinger equation, with formal solution
\begin{equation}
    \ket{\psi(t)} = \exp\{-\rmi \hat{H}_{G} t \}\ket{\psi(0)},
\end{equation}
using units in which $\hbar=1$.

%------------------------------------------------------------------------%
\subsection{Computing using a quantum walk}\label{ssec:qwcomp}
%------------------------------------------------------------------------%

The task is to solve an optimization problem whose $N=2^n$ candidate solutions $j$ are represented in the computational basis $\{\ket{j}\}_{j=0}^{N-1}$, where $j$ is a bit string corresponding to the state of $n$ qubits. 
The problem is encoded in an Ising Hamiltonian $\hat{H}_P$, of the form described by $\hat{H}_I$ in (\ref{eq:Ising_long}) and whose eigenbasis is the computational basis.
We write the basis state with eigenvalue $E^{(P)}_a$ as $\ket{E^{(P)}_a}$, with $a \in \{0\dots N-1\}$, and adopt the convention that $E^{(P)}_a\leq E^{(P)}_{a+1}$.
In other words, $\Big\{\ket{E^{(P)}_a}\Big\}_{a=0}^{N-1}$ is a reordering of $\{\ket{j}\}_{j=0}^{N-1}$ based on the corresponding eigenenergies of $\hat{H}_P$.
The encoding is chosen such that the solution corresponds to the ground state $\ket{E^{(P)}_0}$ of the problem Hamiltonian $\hat{H}_P$.  

To use a quantum walk to solve the problem, we must first choose a suitable state in which to initialize the system. 
With no prior knowledge of the solution, the equal superposition of all basis states
\begin{equation}\label{eq:maxig}
    \ket{\psi(0)}=N^{-1/2}\sum_{j=0}^{N-1}\ket{j},
\end{equation}
is a sensible choice that avoids bias. 
More generally, the initial state can be prepared as weighted or biased superposition, to incorporate prior knowledge about the solution \citep{Perdomo-Ortiz11guessing,Duan2013AQC, chancellor2017modernizing, Grass2017hybrid, Baldwin2018, kechedzhi18a, Grass2019longitudinal}.
Next, we choose a suitable walk graph $G$. 
The main requirement is that the ground state of the quantum walk Hamiltonian $\hat{H}_{G}$ coincides with the initial state, either biased or unbiased (see section \ref{ssec:energycons}). A simple way to achieve a biased starting state would be to `tilt' the driver fields so they are no longer completely transverse.  
We only treat the unbiased case in this work, so our initial state will be $\ket{\psi(0)}$ throughout. 
The full Hamiltonian $\hat{H}(\gamma)$ is defined by adding the quantum walk Hamiltonian $\hat{H}_{G}$ to the problem Hamiltonian $\hat{H}_P$
\begin{equation}\label{eq:Hqwcomp}
    \hat{H}(\gamma) \equiv \hat{H}_{G} + \hat{H}_P \label{eq:fullH},
\end{equation}
where the key parameter is the hopping rate $\gamma$ in $\hat{H}_{G}$, see (\ref{eq:qwcont}).
The computation is performed by evolving the initial state (\ref{eq:maxig}) under the full Hamiltonian $\hat{H}(\gamma)$ for a time $t_f$, then measuring the qubit register in the computational basis. 
The intuition, based on the faster spreading of quantum walks over classical found in prior work \citep{farhi98a}, is that the quantum walk dynamics provide rapid exploration of the basis states, while the energy structure of the problem Hamiltonian $\hat{H}_P$ causes localisation around low-energy states. 

The success probability $P(t_f)=\left|\braket{E^{(P)}_0}{\psi(t_f)}\right|^2$ of finding the solution state when measuring will not in general be unity. 
It will typically be necessary to repeat the protocol multiple times to obtain a high probability of success over all the repeats. 
In general, it will be best to use different measurement times $t_f$ for each repeat. 
Different measurement times will produce different success probabilities $P(t_f)$, and varying the measurement time avoids repeatedly measuring at a time for which the probability $P(t_f)$ happens to be atypically small. 
More precisely, we choose the measurement time $t_f$ uniformly at random in an interval $[t,t+\Delta t]$, and define an \emph{average single run success probability} 
\begin{equation}\label{eq:gensuccprob}
    \bar{P}(t,\Delta t) 
    \equiv \frac{1}{\Delta t}\intop_t^{t+\Delta t}\mathrm{d}t_f P(t_f).
\end{equation}
Operationally, choosing the measurement time $t_f$ randomly in the interval $[t,t+\Delta t]$ samples success probabilities from the distribution with $\bar{P}(t,\Delta t)$ as its mean.
Sampling measurement times in this way means that the protocol typically needs to be repeated $M_\mathrm{rep}\sim 1/\bar{P}(t,\Delta t)$ times to achieve an overall $O(1)$ success probability. 
Note that it is not generally possible to check whether the state measured is indeed the ground state of $\hat{H}_P$. 
However, it is easy to calculate the energy of the state measured in each repeat. 
If only the lowest energy state is accepted, it is only necessary for the ground state of $\hat{H}_P$ to be measured once out of all the repeats.  The more repeats, the more confidence is gained that the lowest energy state
found is the ground state.
And studying the distribution of the sampled energies can provide more information about the problem.

The procedure described in this subsection does not in general provide an optimal quantum algorithm, because the repeats do not use information gained from the outcomes of previous runs.
We will discuss this further in section \ref{sec:outlook}; for most of this paper we are concerned with understanding the average single run success probability, as an essential prerequisite to building optimal algorithms.

In the limit of small interval width $\Delta t$, the average success probability defined in (\ref{eq:gensuccprob}) reduces to the single time probability $P(t_f)=\lim_{\Delta t\rightarrow 0}\bar{P}(t_f,\Delta t)$.
The long time limit of this average,
\begin{equation}\label{eq:infsuccprob}
    P_\infty \equiv \bar{P}(0,\infty) \equiv \lim_{\Delta t\rightarrow \infty}\bar{P}(0,\Delta t),
\end{equation}
is particularly useful, because it can be calculated via a numerical diagonalization of the Hamiltonian (see section \ref{sec:numericalmethods}) and it predicts the short time average well (see subsection \ref{ssec:mixtime}). In this paper, we will often use the long time average $P_\infty$ as an indication of the success probability achievable in a single run, and thus the number of repeats required to achieve $O(1)$ success probability overall. We will separately address the timescale required to reach this probability in each run.

%------------------------------------------------------------------------%
\subsection{Graph choice for quantum walk computing}
%------------------------------------------------------------------------%

There are many graph-based Hamiltonians with the initial state $\ket{\psi(0)}$ defined in (\ref{eq:maxig}) as the ground state. 
A common choice is the \emph{complete graph} $K$, in which every vertex is connected to every other. 
This graph has the quantum walk Hamiltonian $\hat{H}_K$ that couples every computational basis state $\ket{j}$ state to every other,
\begin{eqnarray}\label{eq:KN}
    \hat{H}_K&=\gamma\left[N\mathbb{1}-\sum_{j,k=0}^{N-1}\ketbra{j}{k}\right] \nonumber\\
               &= \gamma N\left[\mathbb{1} - \ketbra{\psi(0)}{\psi(0)}\right].
\end{eqnarray}
The complete graph is useful because it makes some algorithms analytically tractable
\citep[see, e.g.,][]{childs2004spatial}. 
However, for implementation on qubit-based hardware, the complete graph is not in general practical, requiring higher order interaction terms than the transverse Ising term (\ref{eq:Ising_trans}). 
In this qubit setting, an implementation of the complete graph requires a sum over every one-body term (e.g $\hat{X}_j$), every two-body term (e.g $\hat{X}_j\hat{X}_k$), every three-body term (e.g $\hat{X}_j\hat{X}_k\hat{X}_l$) ... up to the $n$-body term $\prod_{j=0}^{n-1}\hat{X}_j$, a total of $N$ terms.
One- and two-body terms are relatively easy to implement, since they correspond to Hamiltonians found naturally.
Terms in three or more Pauli-$X$ operators are much more difficult and generally require extra qubits to engineer in real physical systems.

\begin{figure}
\centering
    \includegraphics[width=0.4\textwidth]{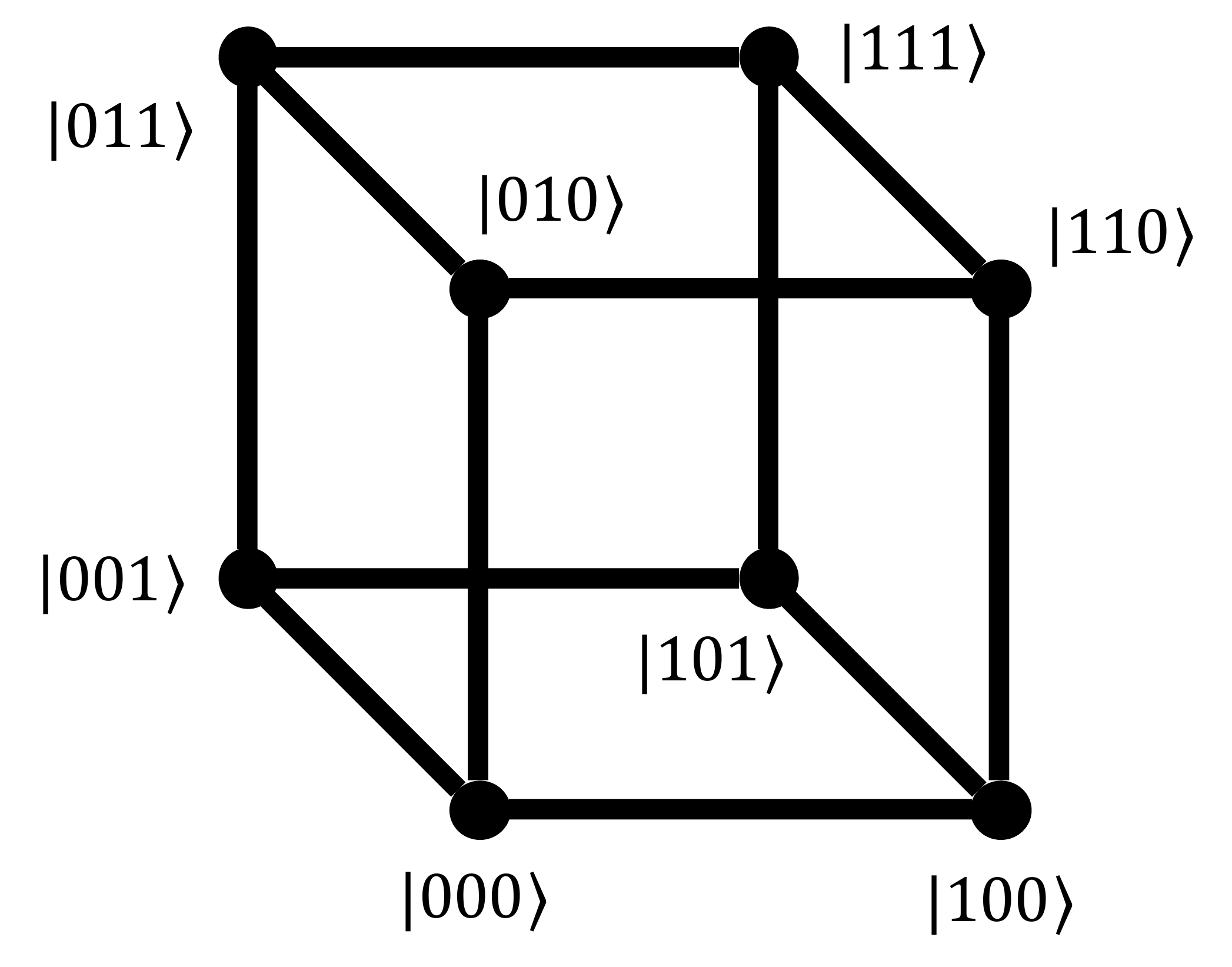}
    \caption{A 3-dimensional hypercube (a cube) graph in which the vertices are labeled by the $2^3=8$ computational basis states  of 3-qubits, and the edges connect the states with Hamming distance 1 (single spin flips).
    \label{fig:hypercube} 
    }
\end{figure}

A more natural choice of graph for qubits is the hypercube. 
The $n$-bit labels are associated with the vertices of the graph such that the edges correspond to flipping one bit, as illustrated in figure \ref{fig:hypercube}. 
The hypercube quantum walk Hamiltonian $\hat{H}_h$ on $n$ qubits is composed of single-body terms
\begin{eqnarray}\label{eq:hypercubehamiltonian}
    \hat{H}_h = \gamma\left[n\mathbb{1}-\sum_{j=0}^{n-1}\hat{X}_j\right].
\end{eqnarray}
With $\hat{H}_h$ as the graph Hamiltonian, the full quantum walk computational Hamiltonian $\hat{H}(\gamma)$ defined in (\ref{eq:fullH}) is a transverse Ising Hamiltonian in the form of $\hat{H}_\mathrm{TI}$ in (\ref{eq:TIM}), with the control functions $A(t)$ and $B(t)$ kept constant throughout the computation.
In this work, we predominantly use the hypercube graph, with some comparisons made with the same problems on the complete graph.

%-----------------------------------------------------------------------------%
\subsection{Solving the search problem using quantum walks}\label{ssec:qsearch}
%-----------------------------------------------------------------------------%

The simplest example of an algorithm in this continuous-time quantum walk setting is the search problem.  
The problem is to find the marked state, a single bit-string $m\in\{0,1\}^n$ out of $N=2^n$ possible bit strings. 
Finding a marked state was shown to have a quantum algorithm with a speed up over classical algorithms by \cite{grover1996fast}. 
To map this problem to the continuous-time 
Hamiltonian setting, the marked basis state $\ket{m}$ is given one less unit of energy than all the rest of the basis states, by defining the problem Hamiltonian $\hat{H}_\mathrm{S}$ as
\begin{eqnarray}
    \hat{H}_\mathrm{S} = -|m\rangle\langle m|.
\end{eqnarray}
By construction, the problem Hamiltonian $\hat{H}_\mathrm{S}$ has the marked state $|m\rangle$ as its ground state. 

The continuous-time quantum walk search problem has been analytically solved 
\citep{childs2004spatial} for several different walk graphs. 
For the complete graph and the hypercube graph, a quantum speed up is obtained for carefully chosen optimal values of the hopping rate $\gamma$.
For the complete graph Hamiltonian $\hat{H}_K$, the optimal value is $\gamma^{(K)}_\mathrm{opt} = 1/N$, while for the hypercube Hamiltonian, $\hat{H}_h$, the optimal hopping rate $\gamma^{(h)}_\mathrm{opt}$ is given by  
\begin{equation}\label{eq:ssgamma}
    2\gamma^{(h)}_\mathrm{opt} = \frac{1}{N}\sum_{r=1}^n \binom{n}{r}\frac{1}{r},
\end{equation}
where $\binom{n}{r}=\frac{n!}{r!(n-r!)}$ is the binomial coefficient. 
For a quantum speed up, the hopping rate must be set to $\gamma^{(h)}_\mathrm{opt}$ as defined by (\ref{eq:ssgamma}) with high precision. 
It has been shown \citep{morley2017quantum} that the fractional tolerance to misspecification of the optimal hopping rate $\gamma^{(h)}_\mathrm{opt}$ falls as $O(N^{-1/2})$. 

The measurement time must also be chosen appropriately. In the limit of large problem size $N$, the marked state can be found with unit success probability, $\lim_{N\rightarrow\infty}\Big[P(t^\mathrm{(opt)}_f)\Big]=1$, by measuring in the computational basis at an optimal measurement time $t^\mathrm{(opt)}_f$.
For both the hypercube and complete graphs, the optimal time $t^\mathrm{(opt)}_f$ scales with the square-root of the problem size $N$ as $t^\mathrm{(opt)}_f\simeq\frac{\pi}{2}N^{1/2}$.
This corresponds to a quadratic speed up compared to the best classical algorithm. Due to the absence of structure in the search problem specifically, such a quadratic speed up has been proven to the best possible quantum speed up \citep{bennett1997strengths}. 

\begin{figure}
  \subfigure[]{\includegraphics[width=0.49\textwidth]{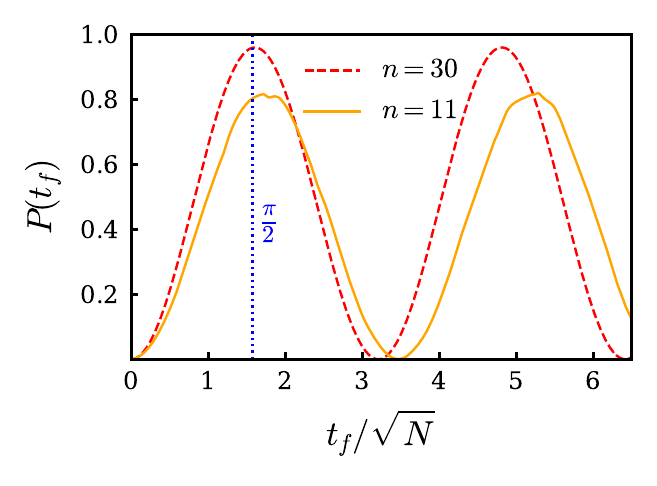}\label{fig:rabiosc}}
  \subfigure[]{\includegraphics[width=0.49\textwidth]{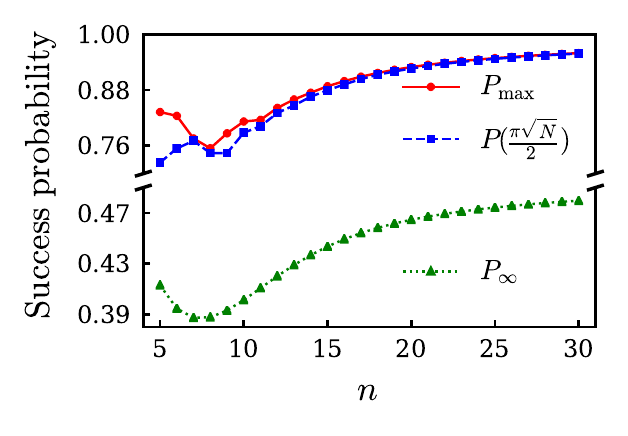}\label{fig:ss_finitesizeeffects}}  
  \caption{The search problem solved using a continuous-time quantum walk on the hypercube using the optimal hopping rate $\gamma^{(h)}_\mathrm{opt}$ given by (\ref{eq:ssgamma}).  
  (a) The probability $P(t_f)$ that a measurement at time $t_f$ results in successfully finding the marked state $|m\rangle$ for two different numbers $n=30$ (red, dashed line) and $n=11$ (orange, solid line) of qubits (i.e problem sizes $N=2^{30}$ and $N=2^{11}$ respectively).
  (b) Comparison of instantaneous success probabilities $P(t_f)$, at the asymptotically optimal (blue squares, dashed line) and numerically determined best (red circles, solid line) measurement times $t_f$, and the infinite time average success probability $P_{\infty}$ (green triangles, dotted line) defined in (\ref{eq:infsuccprob}). 
  \label{fig:ssdynamics}
  }
\end{figure}

The variation of $P(t_f)$ with $t_f$ is shown in figure \ref{fig:rabiosc} for search on hypercube graphs of size $N=2^{30}$ (i.e., $n=30$ qubits) and $N=2^{11}$ (i.e., $n=11$ qubits), using the optimal hopping rate $\gamma^{(h)}_\mathrm{opt}$.
The sinusoidal oscillations of the probability $P(t_f)$ occur because the quantum walk is performing Rabi oscillations between the initial state and the marked state.  The two lowest energy levels of the full Hamiltonian $\hat{H}(\gamma)$ with varying $\gamma$ undergo an avoided level crossing at $\gamma^{(h)}_\mathrm{opt}$ and
the associated eigenstates $\ket{E_{0}(\gamma^{(h)}_\mathrm{opt}})$ and $\ket{E_{1}(\gamma^{(h)}_\mathrm{opt}})$ are approximately the orthogonal equal superpositions of the starting state and marked state, $\ket{E_{0,1}(\gamma^{(h)}_\mathrm{opt}})\simeq(\ket{\psi(0)} \pm \ket{m})/2^{\frac{1}{2}}$. 
The gap $E_1(\gamma^{(h)}_\mathrm{opt})-E_0(\gamma^{(h)}_\mathrm{opt})$ scales with the problem size $N$ as $O(N^{-1/2})$ \citep{childs2004spatial}. 

These simple, two-level dynamics describe the quantum walk solution to the search problem well for large problem size $N$: the oscillations in the $N=2^{30}$ case have no visible irregularities. 
For smaller sizes, finite-size effects due to population of higher energy levels are apparent: the oscillations in the $N=2^{11}$ case have lower probability peaks and show some irregular behaviour, such as the small dip on the first peak. 
These finite-size effects are further illustrated in figure \ref{fig:ss_finitesizeeffects}, which shows the instantaneous success probability $P(t_f)$ at the asymptotically optimal and numerically determined best times, as well as the infinite-time average success probability $P_{\infty}$ defined in (\ref{eq:infsuccprob}).
All three probabilities show a pronounced dip around $n=8$ qubits, with smooth behaviour only settling in for $n> 12$ qubits.
Figure \ref{fig:ss_finitesizeeffects} also shows that the infinite-time probability $P_\infty$ asymptotes to a half. 
Hence, a quantum walk search with a random measurement time should on average only need to be repeated twice to locate the marked state; knowing the exact time to measure for the optimal success probability is not necessary for the success of the algorithm.
Fixed point quantum search algorithms \citep{Yoder2014fixedpoint,Dalzell2017fixedpoint} are another approach that avoids the need to know how long to run the algorithm for.

The search problem in the continuous-time quantum computing setting has two important drawbacks. 
Firstly, implementing the problem Hamiltonian $\hat{H}_S$ directly on $n$ qubits requires $O(2^n)$ terms of products of up to $n$ Pauli-$Z$ operators, similar to the problem with implementing the complete-graph Hamiltonian $\hat{H}_K$, defined in (\ref{eq:KN}), on qubits.
Implementing higher order Pauli-$Z$ terms can be done using extra qubits as ``gadgets'', e.g.,  \citep{Jordan08gadget}.  An alternative type of gadget, specifically for permutation-symmetric problems like search, is given in \citep{dodds18a}, building on classical problem mapping techniques in \citep{chancellor16a,chancellor17a}.
Secondly, it is impossible to map the problem Hamiltonian to qubits without specifying the solution outright.  Hence, the search problem serves as a useful toy problem, especially in contexts where having analytic, computational, and physical implementations available for comparisons facilitates benchmarking and other testbed procedures. 

%%%%%%%%%%%%%%%%%%%%%%%%%%%%%%%%%%%%%%%%%%%%%%%%%%%%%%%%%%%%%%%%%%%%%%%%%%
\section{Spin glass problem definitions}\label{sec:spinglassproblems}
%------------------------------------------------------------------------%

In this work we focus on spin glass problems that have features in common with real life hard optimizations problems and, unlike the search problem, do not admit analytic solutions.  
The search problem solved by quantum walk provides useful comparisons with these spin glass problems.

%------------------------------------------------------------------------%
\subsection{Sherrington-Kirkpatrick spin glass}\label{ssec:sk_introduction}
%------------------------------------------------------------------------%

The Sherrington-Kirkpatrick (SK) spin glass Hamiltonian $H_\mathrm{SK}$ \citep{kirkpatrick1975solvable} is defined on $n$ spins as
\begin{eqnarray}\label{eq:SKorig}
    H_\mathrm{SK} = -\frac{1}{2}\sum_{(j\neq k)=0}^{n-1}J_{jk}S_jS_k
\end{eqnarray}
where $S_j$ are the classical spins ($S_j\in\{-1,1\}$) and the couplings $J_{jk}$ are drawn independently from the normal distribution $\mathcal{N}(\mu,\sigma_\mathrm{SK}^2)$ with mean $\mu$ and variance $\sigma_\mathrm{SK}^2$. 
Finding the ground state of this Hamiltonian is NP-hard \citep{choi2010adiabatic}, and \textit{uniformly} hard, due to its finite-temperature phase transition \citep{kirkpatrick1975solvable}. 

It is computationally convenient to break the spin inversion symmetry by adding single-body field terms of the form $\sum_{j=0}^{n-1}h_{j}S_k$, where $h_j$ are the field strength values. 
Like the couplings $J_{jk}$, the fields $h_j$ are also drawn independently from $\mathcal{N}(\mu,\sigma_\mathrm{SK}^2)$.
When the fields strengths $h_j$ are drawn from the same distribution as the coupling strengths $J_{jk}$, the hardness of finding the ground state follows directly from the hardness of the $h_j=0$ case. 
The SK spin glass with such fields is mathematically equivalent to a zero field spin glass with one more spin which is ``fixed'' in one orientation. 
This is not true in general for different distributions of field strength $h_j$. 
There are known examples in which fields can destroy spin glass behaviour \citep[see, e.g.,][]{Young04a,Feng14a}. 
In particular, if the field strengths are much larger than the coupling strengths ($|h_j| \gg |J_{jk}|$ for all $j,k$), then the energy is minimized trivially when all the spins each minimize the energy with respect to their individual fields. 
While the distribution of field strengths could be used to tune the problem hardness, we do not use it in this way here, and only consider cases where the field and coupling strengths are drawn from the same distribution.

An astute reader will notice that if one effectively un-fixes the spin which corresponds to the fields (thus making all states two fold degenerate and converting the system to a double cover of the orignal system), these couplings will effectively be on average stronger by a factor of $\sqrt{2}$. As this increase in coupling strength does not scale with the number of spins, it is going to become less and less significant as the size of the system is scaled up the hardness will be preserved.

The mapping into the quantum Ising model is almost trivial: the classical spin variables $S_j$ are simply mapped to Pauli-$Z$ operators. 
Thus, the problem Hamiltonian $\hat{H}_\mathrm{SK}$ becomes
\begin{eqnarray}\label{eq:SKh}
    \hat{H}_\mathrm{SK} = -\frac{1}{2}\sum_{(j\neq k)=0}^{n-1}J_{jk}\hat{Z}_j\hat{Z}_k - \sum_{j=0}^{n-1}h_j\hat{Z}_j,
\end{eqnarray}

The SK problem Hamiltonian differs from the search problem by having structure, produced by the $\hat{Z}_j\hat{Z}_k$ terms. 
As a result, the covariances between the energies of two basis states depends on the Hamming-distance between them \citep{Baldwin2018}. 
Knowing the energy of one state gives some information about the energy of states that differ by a small number of bit-flips.  
This results in a distribution of the eigenenergies that is almost normal (as can be seen by plotting the distributions and numerically calculating moments), but which deviates from normal in the tails of the distribution.  

%------------------------------------------------------------------------%
\subsection{Random energy model}\label{ssec:REM}
%------------------------------------------------------------------------%

To isolate the effect of the correlations in the SK problem, we compare it with the random energy model (REM) \citep{derrida1980random}, in which the eigenenergies themselves are independently drawn from a normal distribution. 
The problem Hamiltonian $\hat{H}_\mathrm{REM}$ for REM is 
\begin{eqnarray}\label{eq:rem}
    \hat{H}_\mathrm{REM} &=& \sum_{j=0}^{N-1}F_j\ketbra{j}{j},
\end{eqnarray}
with $\{\ket{j}\}_{j=0}^{N-1}$ the computational ($Z$) basis and the energies $F_j$ drawn independently from the normal distribution $\mathcal{N}(0,\sigma_\mathrm{REM}^2)$. 

REM has a similar energy level distribution to that of SK, apart from the tails.  By definition it lacks the correlations: knowing the energy of one state gives no information about the energies of other states. 
Comparison between these two models highlights the effect of the pairwise structure in the SK model.

%%%%%%%%%%%%%%%%%%%%%%%%%%%%%%%%%%%%%%%%%%%%%%%%%%%%%%%%%%%%%%%%%%%%%%%%%%
\section{Numerical methods}\label{sec:numericalmethods}
%------------------------------------------------------------------------%

The main tool used for the investigations in this work is numerical simulation. 
We are studying computationally hard problems for which there are no tractable analytical solutions except in special cases.

For each number of qubits $5\le n\le 20$ we generated 10,000 random instances of the SK spin glass Hamiltonian, defined in (\ref{eq:SKh}), with the couplings $J_{jk}$ and fields $h_j$ drawn with a standard deviation $\sigma_\mathrm{SK}=\omega_\mathrm{SK}$, where $\omega_\mathrm{SK}$ is an arbitrary energy unit. 
The value $\omega_\mathrm{SK}=5$ was used for computational convenience.
We also generated 10,000 random instances of the REM Hamiltonian, defined in (\ref{eq:rem}), for each number of qubits $5\le n\le 15$, with normally-distributed energies $F_j$ drawn with a standard deviation $\sigma_{\mathrm{REM}}=\omega_{\mathrm{REM}}$. 
The value $\omega_{\mathrm{REM}}=1$ was used for computational convenience.
Note that choosing any arbitrary constant for $\omega$ will only affect overall time and energy scales by a constant factor, and the energy unit $\omega_\mathrm{SK}$ has been scaled out of the plots where relevant. 

The key quantity to determine numerically is the probability that the ground state is found by running a quantum walk computation on each spin glass instance. 
It is particularly convenient to compute the infinite-time probability $P_\infty$ given by (\ref{eq:pinfinity}), for sizes where full diagonalization is possible.
Writing the spectral expansion of the full computational quantum walk Hamiltonian as
\begin{equation}\label{eq:Hqwspec}
    \hat{H}(\gamma) = \sum_{a=0}^{N-1}E_a(\gamma) \ketbra{E_a(\gamma)}{E_a(\gamma)},
\end{equation}
with indices ordered such that $E_a(\gamma) \leq E_{a+1}(\gamma)$ and $\ket{E_a(\gamma)}$ the eigenstate with eigenvalue $E_a(\gamma)$, we can write the instantaneous probability in terms of the spectral expansions as
\begin{eqnarray}
    P(t) &=& \left|\bra{E^{(P)}_0}\exp{(-it\hat{H}(\gamma))}\ket{\psi(0)}\right|^2 \nonumber\\
         &=& \left|\sum_{a=0}^{N-1}\exp{(-itE_a)}\braket{E^{(P)}_0}{E_a(\gamma)}\braket{E_a(\gamma)}{\psi(0)}\right|^2 \nonumber \\
         &=& \sum_{a=0}^{N-1}\left|\braket{E^{(P)}_0}{E_a(\gamma)}\right|^2 
            \left|\braket{E_a(\gamma)}{\psi(0)}\right|^2 \\
         && + \sum_{a\ne b=0}^{N-1}\Bigg[\exp{(-it(E_a-E_b))}\braket{E^{(P)}_0}{E_a(\gamma)} \times \nonumber\\ 
         && \braket{E_a(\gamma)}{\psi(0)}\braket{E_b(\gamma)}{E^{(P)}_0}\braket{\psi(0)}{E_b(\gamma)}\Bigg] \nonumber.
\end{eqnarray}
Assuming no degeneracy (that is, all gaps  $E_a-E_b$ are nonzero), which is justified for the randomized nature of the SK and REM problems, the oscillatory terms cancel in the infinite limit (because $\intop_0^\infty \mathrm{d}t \exp{(-it\theta)}=0$ for nonzero $\theta$) to leave the infinite-time average probability $P_\infty$ given by
\begin{equation}\label{eq:pinfinity}
    P_\infty = \sum_{a=0}^{N-1}\left|\braket{E^{(P)}_0}{E_a(\gamma)}\right|^2 
            \left|\braket{E_a(\gamma)}{\psi(0)}\right|^2.
\end{equation}

All of the numerical simulation in this work has been performed using the Python3 language \citep{van2003python}, aided extensively by the IPython \citep{perez2007ipython} interpreter and the Jupyter Notebook \citep{jupyter} system. 
The numerical heavy-lifting has been done using NumPy \citep{oliphant2006guide}, SciPy \citep{scipy}, and pandas \citep{mckinney2010data}, and the plotting has been done using matplotlib \citep{hunter2007matplotlib}. 
The dynamical simulations have been performed by computing the action of the propagator $\exp{(-it\hat{H}(\gamma))}$ on the initial state $\ket{\psi(0)}$, using the sparse matrix functions within SciPy when possible.  
For the more computationally demanding analyses, we were limited to $n\le 11$ by the computational resources available.
Where relevant, figures in this paper have error bars included. However, in most cases the error bars are much smaller than the size of the marker symbols used and so are not visible. This is due to the size of the data sets (10k instances per value of $n$), which provides a good level of accuracy for the average quantities.

Simulations were run on the Imperial and Durham University high performance computing facilities.
The data for all the instances used is available on a permanent data archive \citep{data_arch}.

%%%%%%%%%%%%%%%%%%%%%%%%%%%%%%%%%%%%%%%%%%%%%%%%%%%%%%%%%%%%%%%%%%%%%%%%%%
\section{Quantum walks with spin glasses}\label{sec:qw_on_sg}
%------------------------------------------------------------------------%

In order to implement a quantum walk algorithm for finding the ground states of the spin glasses defined in section \ref{sec:spinglassproblems}, we follow the procedure described in section \ref{ssec:qwcomp}: 
Choose a quantum walk graph $G$ and associated Hamiltonian $\hat{H}_G$, and add the spin glass Hamiltonian to get the full computational quantum walk Hamiltonian $\hat{H}(\gamma) = \hat{H}_G + \hat{H}_P$, where $\hat{H}_P$ refers to $\hat{H}_\mathrm{SK}$ or $\hat{H}_\mathrm{REM}$ as appropriate.
Since the hypercube is the natural choice of graph for qubit implementations, we use this graph, with quantum walk Hamiltonian $\hat{H}_h$ defined in (\ref{eq:hypercubehamiltonian}), unless otherwise indicated.  
For the initial state $|\psi(0)\rangle$, we use the equal superposition (\ref{eq:maxig}), which is the ground state of the hypercube Hamiltonian $\hat{H}_h$. 

%------------------------------------------------------------------------%
\subsection{Setting the hopping rate}\label{sec:choosinggamma}
%------------------------------------------------------------------------%

In contrast to the search problem, for SK and REM it is impossible to efficiently calculate the optimal hopping rate $\gamma^{(h)}_\mathrm{opt}$ that maximizes the success probability. 
It is not even clear which measure of success probability should be maximized because, unlike the search problem, there will be no efficient way to find the optimal measurement time $t^\mathrm{(opt)}_f$ for any choice of hopping rate $\gamma$.
To bootstrap the investigation, we choose to define the optimal hopping-rate $\gamma^{(h)}_\mathrm{opt}$ with respect to one of the average probabilities defined in (\ref{eq:gensuccprob}); in particular, we choose the hopping rate that maximizes the infinite-time average probability $P_\infty$ defined in (\ref{eq:infsuccprob}). 
We make this choice because the infinite-time average probability $P_\infty$ is numerically convenient to calculate, and because it has been seen to be a relevant measure of probability in the search example, see figure \ref{fig:ss_finitesizeeffects}. We will see in Subsection \ref{ssec:mixtime} that the probability $P_\infty$ typically agrees well with probabilities averaged over shorter and more practical time windows.

\begin{figure}
 \centering
 \includegraphics[width=1\textwidth]{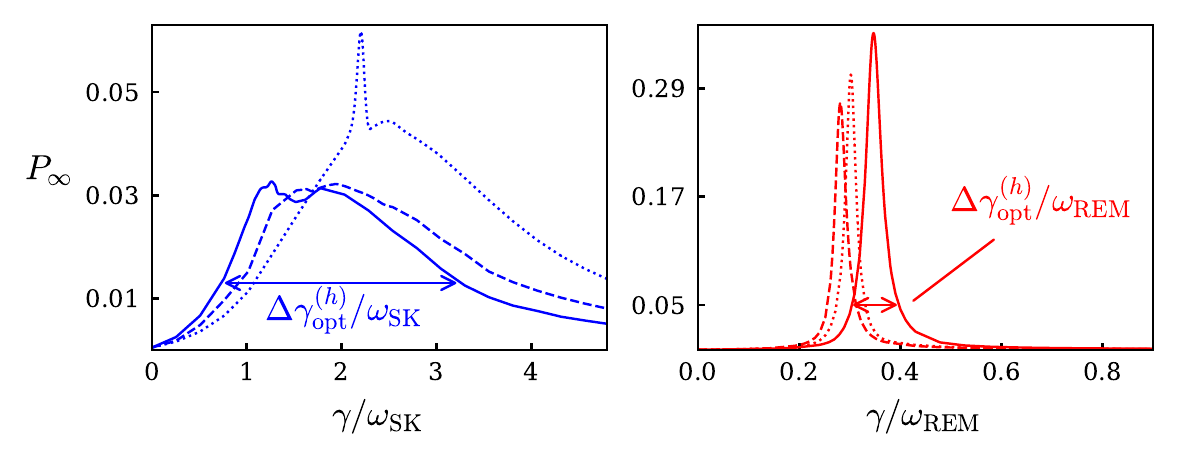}
 \caption{\label{fig:gamma_examples}
 Infinite-time success probability $P_{\infty}$ against hopping rate $\gamma$ scaled by the energy unit $\omega_P$ for 3 typical 11-qubit examples of SK (left) and REM (right). 
 Also indicated (for one example in each plot) is the width $\Delta \gamma^{(h)}_\mathrm{opt}$ of the peak (also scaled by $\omega_P$).
 }
\end{figure}

Some plots of the infinite-time probability $P_\infty$ against hopping rate $\gamma$ for typical 11-qubit examples of the SK and REM are shown in figure \ref{fig:gamma_examples}.
Note that the maximal success probability varies by an order of magnitude between the two problem-types, with REM highest and SK lowest.  
While the optimal hopping rate $\gamma^{(h)}_\mathrm{opt}$ is instance-dependent, these plots show that the dependence of infinite-time probability $P_\infty$ on hopping rate $\gamma$ is typically characterised by broad, bumpy peaks for SK, and by narrow, well-defined peaks for REM. 
This implies that a precise value of the hopping rate $\gamma$ is needed for REM, while there is some tolerance to non-optimal values of the hopping rate $\gamma$ for SK for the sizes that we have studied.

To investigate the success probability more systematically, we performed a brute-force numerical search to find the optimal hopping rate $\gamma^{(h)}_\mathrm{opt}$ that maximizes the success probability $P_\infty$ for each spin glass instance from the data sets of 10k random instances for $5\le n\le 11$. 
This gives a baseline maximum average single run success probability for the quantum walk algorithm.

\begin{figure}
 \centering
 \includegraphics[width=1.00\textwidth]{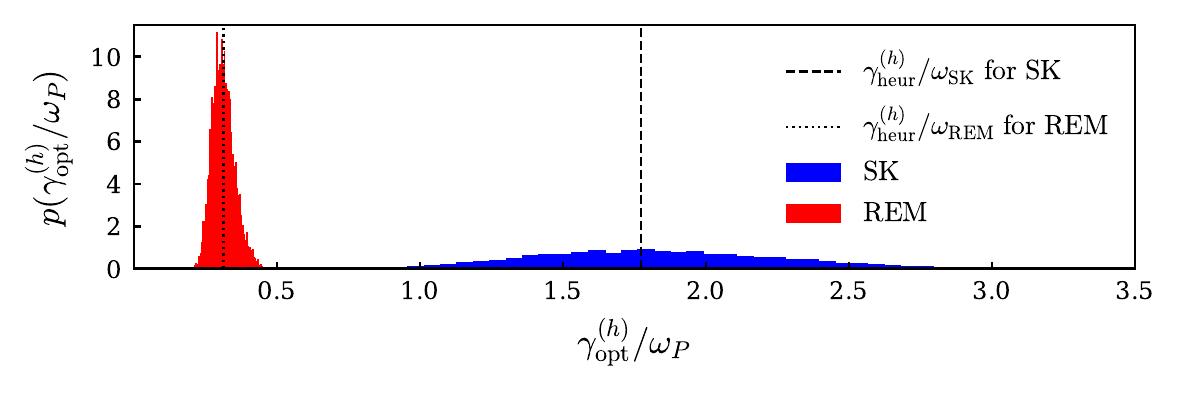}
 \caption{\label{fig:gamma_hists}
 Histograms (relative frequency $p(\gamma^{(h)}_\mathrm{opt}/\omega_P)$) of the numerically-found optimal hopping rates $\gamma^{(h)}_\mathrm{opt}$ scaled by the energy unit $\omega_P$ for the 10,000 11-qubit instances of SK (blue) and REM (red).
 The dashed and dotted lines show the heuristic hopping rate $\gamma^{(h)}_\mathrm{heur}$, calculated according to (\ref{eq:gammaheur}), for SK and REM respectively (also scaled by $\omega_P$).
 }
\end{figure}

The optimal hopping rates $\gamma^{(h)}_\mathrm{opt}$ correspond to the best a quantum walk algorithm on the hypercube can possibly do in a single run. 
For practical algorithms, we need a heuristic method for choosing the hopping rate that can be calculated from the known parameters.
For the quantum walk search algorithm, the optimal hopping rate balances the energy between the two components of the Hamiltonian, $\hat{H}_P$ and $\hat{H}_G$. 
Guided by this, we define the heuristic hopping rate $\gamma^{(h)}_\mathrm{heur}$ for SK and REM such that it balances these overall energy-scales on average.
We match the energy-spread $E_{N-1}^{(h)} - E_{0}^{(h)}$ of the hypercube quantum walk Hamiltonian $\hat{H}_h$ with the \emph{average} energy-spread $\langle E_{N-1}^{(P)} - E_{0}^{(P)} \rangle$ of the problem Hamiltonian $\hat{H}_P$. 
For the hypercube Hamiltonian $\hat{H}_h$ defined in (\ref{eq:hypercubehamiltonian}), we have the energy spread $E_{N-1}^{(h)} - E_{0}^{(h)} = 2n\gamma$; hence, we define the heuristic hopping rate $\gamma^{(h)}_\mathrm{heur}$ by
\begin{equation}\label{eq:gammaheur}
    \gamma^{(h)}_\mathrm{heur} \equiv \frac{1}{2n}\left\langle E_{N-1}^{(P)}-E_{0}^{(P)} \right\rangle.
\end{equation} 
To demonstrate that this heuristic is sensible, we compare in figure \ref{fig:gamma_hists} the distributions of optimal hopping rates $\gamma^{(h)}_\mathrm{opt}$ for SK (blue) and REM (red), as well as the heuristic hopping rates (black, dashed and dotted lines for SK and REM respectively) calculated according to (\ref{eq:gammaheur}), for the 11-qubit data set.
For both SK and REM, the heuristic hopping rate $\gamma^{(h)}_\mathrm{heur}$ falls in the centre of the $\gamma^{(h)}_\mathrm{opt}$ distributions.
Note that the SK distribution is much broader than for REM: not only are the individual peaks for $\gamma^{(h)}_\mathrm{opt}$ for SK much broader than for REM (figure \ref{fig:gamma_examples}), but the distribution of the maxima of those peaks is also much broader (figure \ref{fig:gamma_hists}).  This may seem to be a problem for specifying a heuristic value for $\gamma$ for SK from average energies, but as we will show, it is actually REM that fails for the heuristic $\gamma$, while SK works well.

For a normal distribution of energy levels, the average problem energy spread can be estimated as
\begin{equation}\label{eq:heur_es}
    \left\langle E_{N-1}^{(P)}-E_{0}^{(P)}\right\rangle
    \simeq -(2^{\frac{3}{2}}\sigma_P^{(\mathrm{energy})})\,\mathrm{erf}^{-1}\Big(\frac{1}{N}-1\Big),
\end{equation} 
where $\sigma_P^{(\mathrm{energy})}$ is the standard deviation of the energy eigenvalues of the problem Hamiltonian. 
For REM, the standard deviation $\sigma_\mathrm{REM}^{(\mathrm{energy})}$ is equal to the energy unit $\omega_\mathrm{REM}$ by definition (see section \ref{ssec:REM}). 
For SK, the standard deviation $\sigma_\mathrm{SK}^{(\mathrm{energy})}$ can be shown to be equal to $\frac{\omega_\mathrm{SK}}{2}[n(n+3)]^{\frac{1}{2}}$.
Equation (\ref{eq:heur_es}) is accurate for REM (which has normally-distributed energy levels by definition) but, as already noted, the distribution of the eigenenergies in SK deviates from normal, especially in the tails.  Numerically, we find that there is a multiplicative constant factor of approximately 0.887 that corrects the formula in (\ref{eq:heur_es}) for SK for the effects of the non-normal tails. 
For the numerical analysis, we use the numerically calculated average energy-spread at each number of qubits $n$.

\begin{figure}
  \centering
  \subfigure[]{\includegraphics[width=0.49\textwidth]{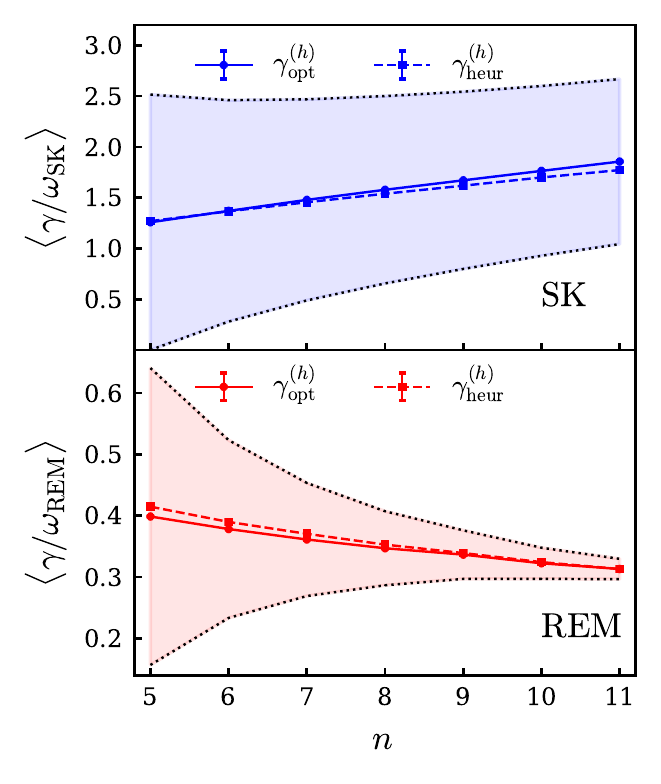}\label{fig:gammacompare}}
  \subfigure[]{\includegraphics[width=0.49\textwidth]{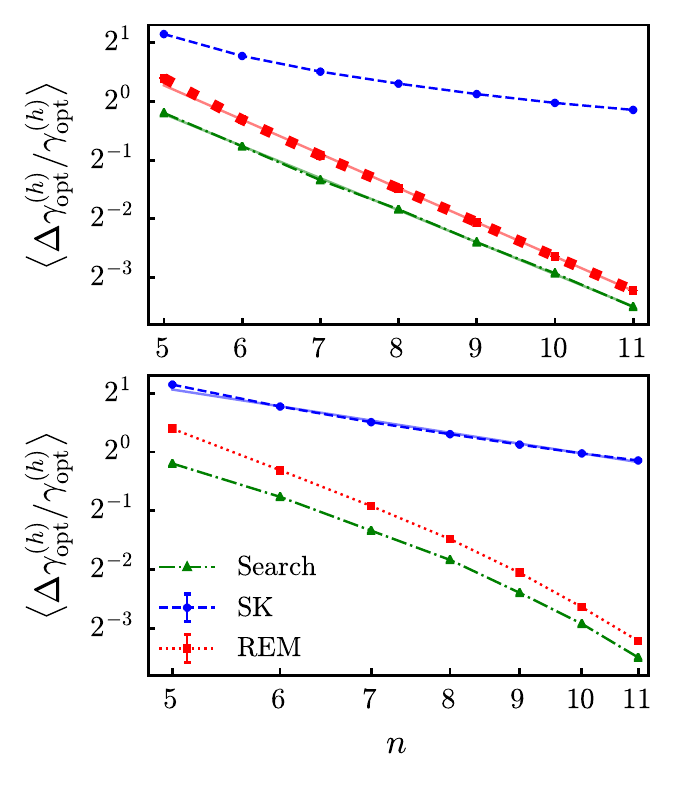}\label{fig:gammawidth}}
  \caption{(a) Average optimal (circles, solid line) and heuristic  (squares, dashed line) hopping rates, $\langle\gamma^{(h)}_\mathrm{opt}\rangle$ and $\gamma^{(h)}_\mathrm{heur}$, against number $n$ of qubits for SK (top, blue) and REM (bottom, red). 
 The shaded regions bordered by dotted lines indicate the average tolerance range $\langle\Delta\gamma^{(h)}_\mathrm{opt}\rangle$ to non-optimal hopping rates, defined as full width at half maximum (FWHM) of the probability peak surrounding $\gamma^{(h)}_\mathrm{opt}$, as illustrated in figure \ref{fig:gamma_examples}. 
   (b) Log-linear plot (top) and log-log plot (bottom) of average \emph{fractional} tolerance range $\langle\Delta\gamma^{(h)}_\mathrm{opt}/\gamma^{(h)}_\mathrm{opt}\rangle$. 
 REM (red squares, dotted line) shows an exponential decrease, fitting to a line (red, solid line) with a gradient of $-0.583\pm0.006$ in the log-linear plot. 
 SK (blue circles, dotted line) shows a polynomial decrease, fitting to a line (solid blue) in the log-log plot with a gradient of $-1.09\pm0.04$. 
 The same quantity for the search problem calculated the same way is also shown (green triangles, dash-dotted line), and it fits well to a line (solid green) in the log-linear plot with a gradient of $-0.546\pm0.004$.}
\end{figure}

Figure \ref{fig:gammacompare} compares the heuristic hopping rate $\gamma^{(h)}_\mathrm{heur}$ and average optimal hopping rate $\langle\gamma^{(h)}_\mathrm{opt}\rangle$ at different numbers of qubits $5\le n\le 11$. 
The full width at half maximum (FWHM) has also been calculated for each instance, to estimate the tolerance $\Delta\gamma^{(h)}_\mathrm{opt}$ to deviations from the optimal hopping rate $\gamma^{(h)}_\mathrm{opt}$ (illustrated in figure \ref{fig:gamma_examples}).
The width of the shaded regions in figure \ref{fig:gammacompare} corresponds to the average tolerance range $\langle \Delta\gamma^{(h)}_\mathrm{opt} \rangle$ at each number $n$ of qubits. 
While the heuristic hopping rate differs slightly from the the average optimal hopping rate for SK, the average tolerance range $\langle\Delta\gamma^{(h)}_\mathrm{opt}\rangle$ is much broader, and does not shrink with increasing number of qubits $n$.   
For REM, however, while we see close agreement on average, the tolerance range shrinks quickly with the number of qubits $n$ as the peaks (as in figure \ref{fig:gamma_examples}, right) become narrower. 
This means that the heuristic hopping rate $\gamma^{(h)}_\mathrm{heur}$ is more likely to lie further than $2\Delta\gamma^{(h)}_\mathrm{opt}$ outside of the actual probability peak for each instance, even though it agrees well with the average optimal hopping rate $\langle\gamma^{(h)}_\mathrm{opt}\rangle$.  
Consequently, a quantum walk with the heuristic hopping rate $\gamma^{(h)}_\mathrm{heur}$ does not perform well for most REM instances.

It is instructive to quantify this sensitivity to deviations from the optimal hopping rate $\gamma^{(h)}_\mathrm{opt}$. 
Figure \ref{fig:gammawidth} shows log-linear and log-log plots of the average \emph{fractional} tolerance range $\langle\Delta\gamma^{(h)}_\mathrm{opt}/\gamma^{(h)}_\mathrm{opt}\rangle$ against number $n$ of qubits for SK (blue circles), REM (red squares) and search (green triangles) on the hypercube.
For SK, the fractional tolerance range $\langle\Delta\gamma^{(h)}_\mathrm{opt}/\gamma^{(h)}_\mathrm{opt}\rangle$ decreases as approximately $1/n$, while for REM and search the decrease is approximately $N^{-0.5}$. 
This decrease is expected theoretically for search \citep{childs2004spatial}. 
The fitted lines do not show exactly a square-root dependence (exponent of $-0.5$) due to the finite size effects for small numbers of qubits $n\le 12$.

Thus, we see that REM behaves like the search problem in a quantum walk setting. 
For a precisely optimal hopping rate $\gamma^{(h)}_\mathrm{opt}$, the success probability is high, but this instance-dependent hopping rate is hard to predict, unlike for the analytically tractable quantum walk search algorithm. 
Without this precise hopping rate, quantum walks perform no better than 
guessing for the search problem and for REM.
In contrast, quantum walks applied to SK give a better-than-guessing success probability $P_\infty>1/N$ for the heuristic hopping rate $\gamma^{(h)}_\mathrm{heur}$ calculated according to (\ref{eq:gammaheur}). 

With the conditions under which we can achieve a better-than-guessing success probability characterised for the three problem types, SK, REM, and search, we turn to the scaling of this success probability with problem size $N$.

%------------------------------------------------------------------------%
\subsection{Success probability}\label{ssec:succprobres}
%------------------------------------------------------------------------%

\begin{figure}
  \centering
  \subfigure[]{\includegraphics[width=0.49\textwidth]{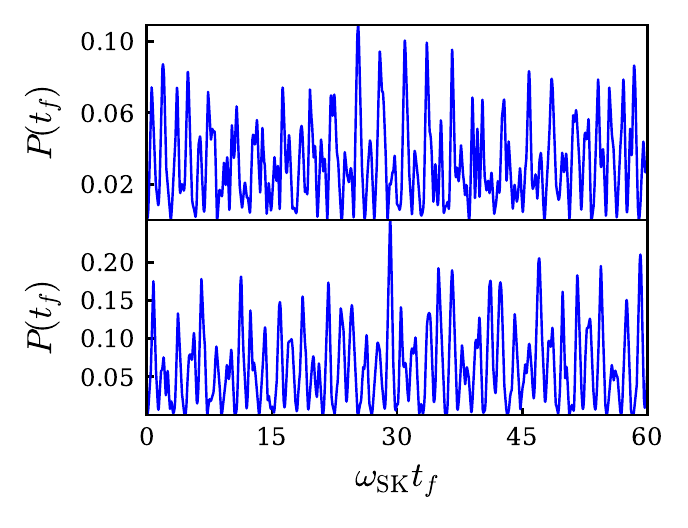}\label{skwalk_examples}}
  \subfigure[]{\includegraphics[width=0.49\textwidth]{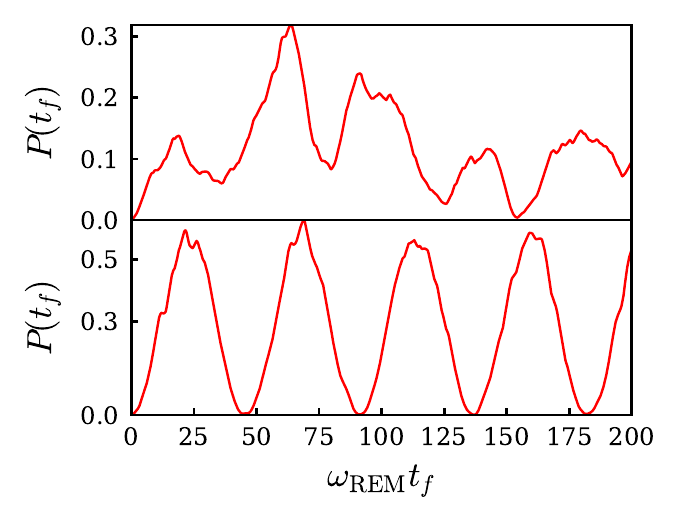}\label{remwalk_examples}}
  \caption{\label{skremwalk_examples}
  Instantaneous success probability $P(t_f)$ against dimensionless measurement time $\omega_P t_f$ for quantum walk on 2 typical 11-qubit SK examples (a) and for 2 typical 11-qubit REM examples (b), using $\gamma^{(h)}_\mathrm{opt}$.
  }
\end{figure}

Figure \ref{skremwalk_examples} shows how the single-time success probability $P(t_f)$ varies with the measurement time $t_f$ for two typical 11-qubit examples of SK and REM. 
In the REM case, the behaviour is similar to that shown in figure \ref{fig:rabiosc} for search: an oscillatory nature indicating the dominance of a two-level avoided-crossing feature, but with evidence of the population of other energy-levels that lead to finite-size effects in search.
For REM, these finite-size effects are more pronounced, and are instance-dependent.  
The random nature of the REM problems means there is not such a clear cut off size, as there is for the search problem, above which finite size effects are negligible.  
In any case, based on search, we expect finite size effects to be significant at $n=11$.
For SK, the behaviour is quite different from search or REM.  
There is no indication of dominant oscillatory behaviour; instead, these plots show unpredictable, highly instance-dependent fluctuating dynamics for all the sizes we are using.
This indicates that for SK, the behaviour is determined by the excitation of many energy levels.

As with finding a suitable hopping-rate $\gamma$, both REM and SK differ from the search problem in that there is no practical way to find the optimal measurement time $t^\mathrm{(opt)}_f$; a different approach must be taken instead.
As already noted for the search problem, this can be handled by using the time averaged probabilities defined in (\ref{eq:gensuccprob}). 
We first consider the infinite-time probability $P_\infty$, as defined in (\ref{eq:infsuccprob}), since it is easy to calculate (see section \ref{sec:numericalmethods}).
\begin{figure}
 \centering
 \includegraphics[width=1.0\textwidth]{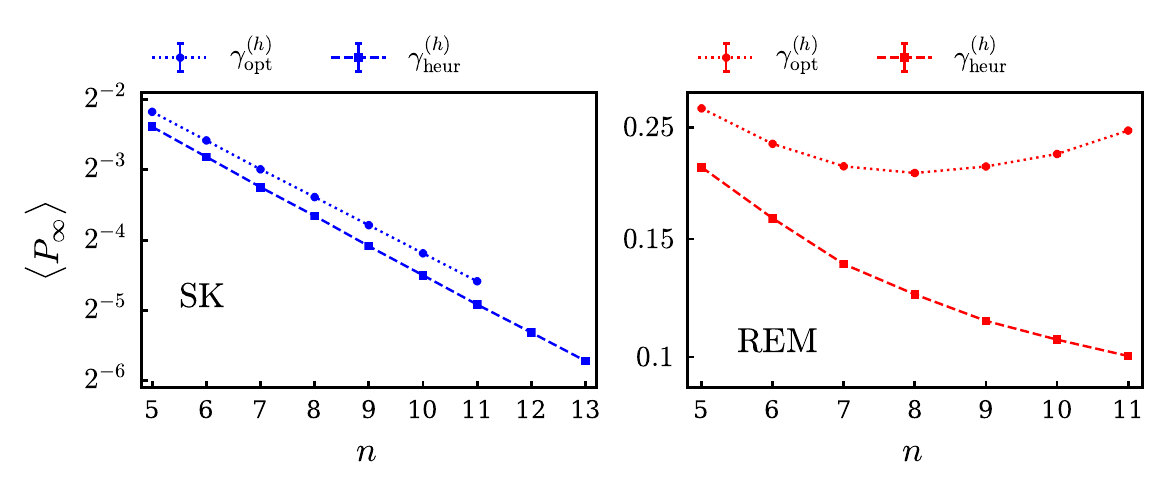}
 \caption{\label{infprobs}
 Blue, left: Log-linear plot of average infinite time success probability $\langle P_\infty\rangle$ against number of qubits $n$ for SK, using optimal (circles, dotted line) and heuristic (squares, dashed line) hopping rates $\gamma^{(h)}_\mathrm{opt}$ and $\gamma^{(h)}_\mathrm{heur}$. The data fit $\log_2\langle P_\infty \rangle=(-0.402\pm0.001)n+(-0.174\pm0.008)$ and $\log_2\langle P_\infty \rangle=(-0.417\pm0.002)n+(-0.32\pm0.01)$ respectively. 
Red, right: Log-linear plot of the same quantities for REM. In this case, the probability stays at constant order for the optimal rate and decays for the heuristic rate.}
\end{figure}
Figure \ref{infprobs} shows the average infinite-time success probability $\langle P_\infty \rangle$ against the number $n$ of qubits for the two problems using both the optimal $\gamma^{(h)}_{\mathrm{opt}}$ and heuristic $\gamma^{(h)}_{\mathrm{heur}}$ hopping rates. 
For SK, this gives exponential decay with the number of qubits $n$ in both cases: the average probability $\langle P_\infty \rangle$ changes with $n$ according to 
\begin{equation}
\label{p_eq}
\langle P_\infty \rangle = \begin{array}{ll}
\tilde{O}(N^{-0.402\pm0.001}) & \mbox{with }\gamma^{(h)}_\mathrm{opt} \\ 
\tilde{O}(N^{-0.417\pm0.002}) & \mbox{with }\gamma^{(h)}_\mathrm{heur}
\end{array},
\end{equation}
where $\tilde{O}$ may neglect factors logarithmic in its argument. 
That is, using the heuristic hopping rate $\gamma^{(h)}_\mathrm{heur}$ instead of the optimal hopping rate $\gamma^{(h)}_\mathrm{opt}$ has only a minor impact on the average success probability $\langle P_\infty\rangle$. 

For REM, the behaviour is quite different. With the optimal hopping rate $\gamma^{(h)}_\mathrm{opt}$ we see a success probability $P_\infty$ of constant order but with a pronounced dip. 
This behaviour is similar to that seen for the search problem, where the dip seen in figure \ref{fig:ss_finitesizeeffects} is a finite-size effect. 
This similarity is expected, given the similarity between the dynamical behaviour shown in figure \ref{fig:rabiosc} for search and in figure \ref{remwalk_examples} for REM. 
With the heuristic hopping rate $\gamma^{(h)}_\mathrm{heur}$ for REM, we see a significantly reduced success probability $P_\infty$ compared to the optimal case. 
That is, the heuristic is performing poorly, despite the good agreement shown in figure \ref{fig:gammacompare}.

The clear difference in behaviour between SK and REM can be explained by the different tolerances $\Delta\gamma^{(h)}_\mathrm{opt}$ to deviations from the optimal hopping rate $\gamma^{(h)}_\mathrm{opt}$  shown in figure \ref{fig:gammacompare} and figure \ref{fig:gammawidth}.  For SK, the tolerance range is broad enough for the heuristic to lie within it, while for REM the heuristic hopping rate $\gamma_\mathrm{heur}^{(h)}$ almost always misses this range entirely even though it is close to the average optimal hopping rate $\left\langle\gamma_\mathrm{opt}^{(h)}\right\rangle$.

%------------------------------------------------------------------------%
\subsection{Mixing times}\label{ssec:mixtime}
%------------------------------------------------------------------------%

We have thus numerically determined an average success probability scaling with problem size of ${\sim \tilde{O}(N^{-0.42})}$ for a quantum walk finding SK spin glass ground states, using the heuristic hopping rate $\gamma^{(h)}_{\mathrm{heur}}$.
This is based on the infinite time-success probability $P_\infty$, i.e., uniform sampling from the distribution of all possible run times. 
We now investigate the time dependence in more detail: can we sample from a finite run time and still obtain the same speed up?  
Since $P(0)=1/N$ corresponds to random guessing, there must be a minimum time before which it is not effective to measure.

\begin{figure}
  \centering
  \subfigure[]{\includegraphics[width=0.49\textwidth]{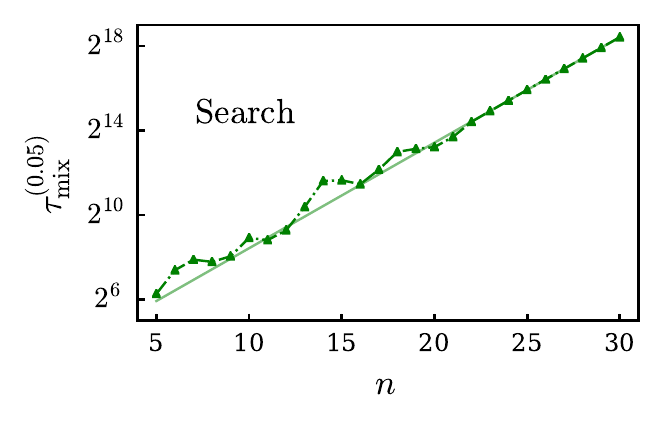}\label{ss_mt}}
  \subfigure[]{\includegraphics[width=0.49\textwidth]{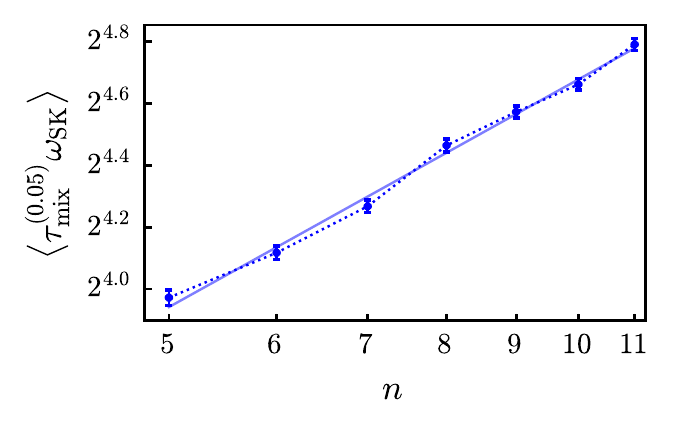}\label{mt}}
  \caption{\label{mt_w_ss}
  (a) Log-linear plot of the mixing time $\tau_\mathrm{mix}^{(0.05)}$ for search, using $\gamma^{(h)}_\mathrm{opt}$.
  The solid line of best fit is $\log_2\tau_\mathrm{mix}^{(0.05)}=(0.5000\pm0.0002)n+(3.424\pm0.006)$, with finite-size effects dominating at small numbers of qubits $n\lesssim20$.
  (b) Log-log plot of the average mixing time scaled by $\omega_\mathrm{SK}$ to give a dimensionless quantity $\langle \tau_\mathrm{mix}^{(0.05)}\omega_\mathrm{SK} \rangle$ against system-size $n$ for SK, using $\gamma^{(h)}_\mathrm{opt}$. 
  The solid line of best fit is $\log_2\langle \tau_\mathrm{mix}^{(0.05)}\omega_\mathrm{SK} \rangle=(0.74\pm0.03)\log_2n+(2.23\pm0.08)$.}
\end{figure}

We define a \emph{mixing-time} $\tau_\mathrm{mix}^{(\epsilon)}$ to be the latest time, $t$, for which the time averaged probabilities $\bar{P}(0,t)$ and $\bar{P}(0,2t)$ at the two times $t$ and $2t$ differ by a fraction greater than the fluctuation parameter $\epsilon$, 
\begin{eqnarray}\label{eq:taumix}
\tau_\mathrm{mix}^{(\epsilon)} &=& \max\{t :\Big|\frac{\bar{P}(0, t)-\bar{P}(0,2 t)}{\bar{P}(0,t)}\Big|>\epsilon\}.
\end{eqnarray}
This definition of $\tau_\mathrm{mix}^{(\epsilon)}$ is based on similar definitions found in prior work \citep{aharonov00a}, with modifications for computational convenience. 
We numerically estimated the mixing-time $\tau_\mathrm{mix}^{(0.05)}$ for each
SK instance up to $n=11$ qubits, using the optimal hopping rate $\gamma^{(h)}_\mathrm{opt}$ for each instance.
We simulated the quantum walk computation dynamics for a successively-doubling duration until a time at which the condition is met was reached. 
The fluctuation parameter $\epsilon=0.05$ corresponds to a deviation of $5\%$.
To verify that the mixing-time $\tau_\mathrm{mix}^{(0.05)}$ correctly captures the relevant dynamical timescale, we also numerically estimated it for the search problem at each system size from $n=5$ to $n=30$ qubits.
The search problem using continuous-time quantum walks can be mapped to the symmetric subspace, allowing larger sizes to be analysed.
The mixing-time $\tau_\mathrm{mix}^{(0.05)}$ for search exhibits the expected exponential timescale: the solid green line of best fit in figure \ref{ss_mt} has the expected scaling with problem size $N$ of $\tau_\mathrm{mix}^{(0.05)}=\tilde{O}(N^{1/2})$.

For search, the scaling is dominated by the run time, the success probability is $O(1)$.
However, this behaviour only emerges clearly above $n\sim20$. 
Below this, the behaviour is influenced by the finite-size effects that arise due to population of higher energy levels.
This means it is not useful to analyse the behaviour of the REM time scaling, finite size effects mask the scaling behaviour for computationally tractable sizes.
However, unlike search and REM, the SK behaviour is influenced by higher energy levels at all sizes, through the frustration provided by the random couplings between the spins. 
Hence, we do not expect to see such finite-size effects in SK; the behaviour is already dominated by the frustration at small sizes. 
Figure \ref{mt} shows a log-log plot of the mixing-time (scaled by $\omega_\mathrm{SK}$) averaged over the ensemble $\langle \tau_\mathrm{mix}^{(0.05)}\omega_\mathrm{SK} \rangle$. 
The solid blue line of best fit has a logarithmic scaling with problem size $N$ of
\begin{eqnarray}
\langle \tau_\mathrm{mix}^{(0.05)}\omega_\mathrm{SK} \rangle &=& O(n^{0.74\pm0.03})
\simeq O([\log_2N]^{0.75}).
\end{eqnarray}
Thus it contributes a logarithmic factor to the overall scaling. 
We emphasise that while this single-run timescale is polynomial in the number of spins $n$, the overall timescale is still exponential in $n$ due to the exponential number of repeats required to achieve $O(1)$ success probability.
\begin{figure}
 \centering
 \includegraphics[width=0.7\textwidth]{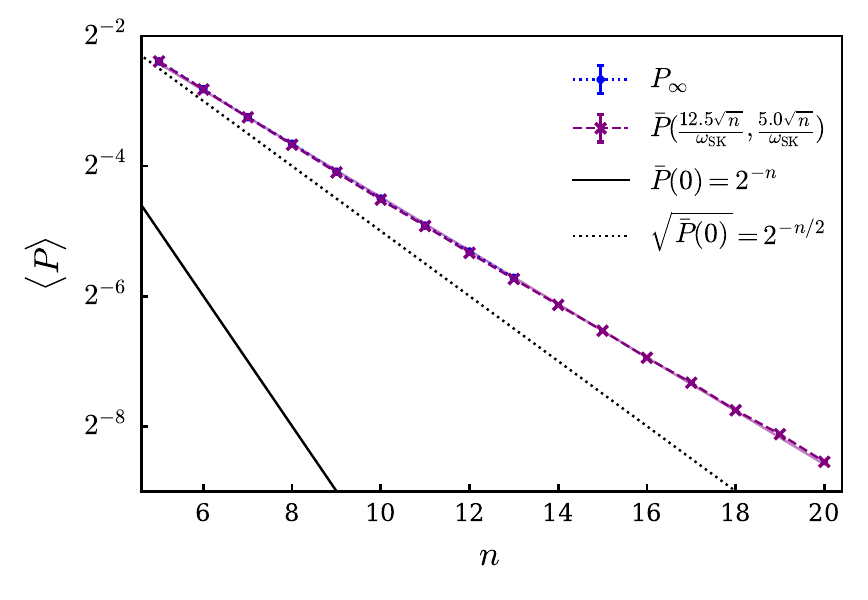}
 \caption{\label{skhf}
 Log-plot of average success probability using $\gamma^{(h)}_{\mathrm{heur}}$ against number of qubits for infinite-time (blue circles and dotted line) as in figure \ref{infprobs}, and averaged over the short time window $12.5n^{\frac{1}{2}}\le t\omega_\mathrm{SK}\le 17.5n^{\frac{1}{2}}$ (purple crosses and dashed line). The short time data are fit by $\log_2\langle\bar{P} \rangle=(-0.410\pm0.002)n+(-0.37\pm0.02)$ (solid purple line). The $2^{-n}$ probability when measuring at $t=0$, equivalent to randomly guessing (solid black line), and its square-root $2^{-n/2}$ (dotted black line) are also shown for comparison.
} 
\end{figure}

To confirm the subsidiary nature of the time scaling for each SK run, we show in figure \ref{skhf} a log-plot comparing, for the heuristic hopping rate $\gamma^{(h)}_{\mathrm{heur}}$, the success probability $P_\infty$ in the infinite-time case (as in figure \ref{infprobs}) and in the case of an early, logarithmically-scaling (with respect to $N$) measurement window $12.5n^{\frac{1}{2}}/\omega_\mathrm{SK} \leq t \leq 17.5n^{\frac{1}{2}}/\omega_\mathrm{SK}\equiv (t_{\mathrm{short}},\Delta t_{\mathrm{short}})$. 
This $n^{0.5}$ scaling of the window is even shorter than the fitted scaling of
$n^{0.75}$, although at these sizes the difference is not significant.
This finite-time probability $\bar{P}(t_{\mathrm{short}},\Delta t_{\mathrm{short}})$
is similar to the infinite-time probability $P_\infty$: the solid purple line of best fit in figure \ref{skhf} has a scaling with problem size $N$ of
\begin{eqnarray}
\label{skhf_peq}
\Big\langle\bar{P}(t_{\mathrm{short}},\Delta t_{\mathrm{short}})\Big\rangle&=&\tilde{O}(N^{-0.410\pm0.002}).
\end{eqnarray}
This should be compared with (\ref{p_eq}), where the value of the exponent for the average infinite time success probability $P_\infty$ with the heuristic hopping rate $\gamma^{(h)}_\mathrm{heur}$ is given by $-0.417\pm0.002$.

As the dominant factor in the total runtime comes from the required number of repeats, and because the single-run timescale contributes only a logarithmic factor, these results constitute good numerical evidence for an average total runtime which scales with problem size $N$ as ${\sim \tilde{O}(N^{0.41})}$ for using quantum walks to find spin glass ground states, over the range of $N$ in our data sets. 
This scaling is a better than the best possible (quadratic) speed up achievable for quantum walk search algorithms.
Moreover, it comes without the requirement for exponential precision in setting the hopping rate that renders practical use of quantum walk searching difficult for large problems.
We now present some insights into where the improvement over search comes from.

%%%%%%%%%%%%%%%%%%%%%%%%%%%%%%%%%%%%%%%%%%%%%%%%%%%%%%%%%%%%%%%%%%%%%%%%%%
\section{Computational mechanisms}\label{sec:compmech}
%------------------------------------------------------------------------%

%-----------------------------------------------------------------%
\subsection{Role of correlations in SK}
\label{ssec:energycorrinsk}
%-----------------------------------------------------------------%

To investigate whether the energy correlations with Hamming distance in SK play a significant role in the computational process of finding the ground state with a quantum walk, we performed three additional sets of numerical tests.

Firstly, we used the same SK instances but performed the quantum walk using a complete graph Hamiltonian $\hat{H}_K$, defined in (\ref{eq:KN}), instead of the hypercube graph Hamiltonian $\hat{H}_h$. 
This removes the correspondence of Hamming-distance between classical states with the distance between those states on the graph -- for the complete graph, every state is one unit (edge) away from every other state. 
In terms of the Hamiltonian, the transverse Ising term is replaced by sums of products of up to $n$ Pauli-$X$ operators that flip up to $n$ qubits at the same time, in all possible combinations. 
For each SK instance up to $n=11$, we estimated the optimal hopping rate $\gamma^{(K)}_\mathrm{opt}$ for the complete graph, and then used it to calculate the infinite-time probability $P_\infty$.

Secondly, we constructed `scrambled SK' instances, denoted sSK, by randomizing which state corresponds to which energy in the SK instances.
In doing so, we arrive at Hamiltonians with identical energy spectra to the SK instances, but without the correlations between energy difference and Hamming distance on the hypercube graph. This approach has similarities with previous work \citep{farhi2008make,farhi2011unstructured,Hen2014Continuous}. 
For each sSK instance, we estimated the optimal hopping rate $\gamma^{(h)}_\mathrm{opt}$, which is different from that used for the ordinary SK versions.  This hopping rate was then used to calculate $P_\infty$.

Thirdly, we sorted the eigenenergies of each REM instance in increasing size and assigned them to the computational basis states in the order of a binary-reflected Gray code on their bitstrings, to arrive at a problem denoted REMGC.
In doing so, we added some amount of Hamming-distance structure by ensuring that the closest energies are assigned to states that differ by only a single bit-flip. 
For each REMGC instance, we estimated an optimal hopping rate $\gamma^{(h)}_\mathrm{opt}$, which is different from that used for the ordinary REM problem.  
This was used to calculate the infinite-time probability $P_\infty$.
While REMGC is not a hard problem as defined, it provides a useful example to compare with how the quantum walk finds the ground state of a SK spin glass.

\begin{figure}
 \centering
 \includegraphics[width=0.7\textwidth]{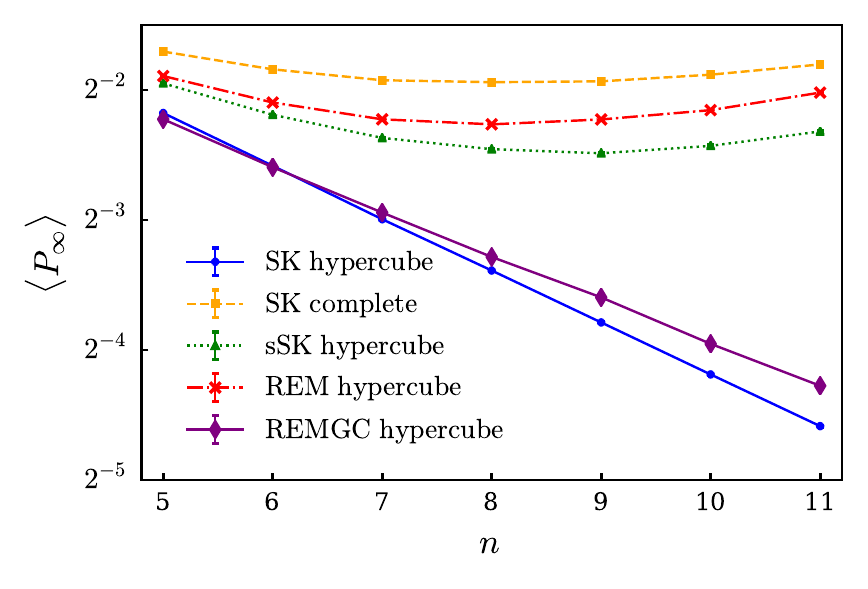}
 \caption{\label{skvariant_compare} 
 Log-linear plot showing the dependence on number of qubits $n$ of the average success probability $P_\infty$ for SK on hypercube (blue circles, thick solid line), REM on hypercube (red crosses, dash-dotted line), sSK on hypercube (green triangles, dotted line), SK on complete-graph (orange squares, dashed line) and REMGC on hypercube (purple diamonds, thin solid line).
 The optimal hopping rates $\gamma^{(h)}_\mathrm{opt}$ are used in all cases.
 }
\end{figure}

These three variants provide separate tests of the influence of the graph structure (choice of quantum walk Hamiltonian) and problem structure (pairwise correlations in SK).
Figure \ref{skvariant_compare} shows how the infinite-time probability $P_\infty$ varies with the number of qubits $n$ for these three variants, alongside SK and REM on a hypercube graph from figure \ref{infprobs}. 
The variation of $P_\infty$ with the number of qubits $n$ for the five variants is clearly split into two groups, behaviour like REM and search on the one hand, and behaviour like SK on the other. 
Removing the correlations from SK by scrambling the energies (sSK) results in behaviour like REM and search. 
Moreover, removing the correspondence between distance and Hamming weight by using the complete graph instead of the hypercube also changes the SK problem behaviour to be like REM and search. 
In the opposite direction, inserting pairwise correlations into REM via a Gray code (REMGC) results in problems that are much more like SK than like the REM problems on a hypercube graph. 

From this, we infer that the problem structure -- in this case the pairwise correlations in SK --  needs to be matched by a compatible driver Hamiltonian -- in this case the hypercube/transverse Ising -- to obtain better than quadratic scaling. 
This type of local structure in the solution space is exploited in many classical algorithms. 
For example, classical Monte Carlo optimizations that use a single bit flip update rule are naturally using this hypercube structure. 
Using a complete graph instead would correspond to flipping a random number of bits, which is equivalent to guessing at each step.

%------------------------------------------------------------------------%
\subsection{Energy conservation dynamics}\label{ssec:energycons}
%------------------------------------------------------------------------%

Continuous-time quantum walk time evolution is unitary, and there is no time dependence in the Hamiltonian that can lead to energy gain or loss by the system.
Hence, it is important to consider how it can find a lower energy state than it starts in (with respect to $\hat{H}_P$) with any better-than-guessing probability. 
For the search problem, this happens through an analog of Rabi flopping (see figure \ref{fig:ssdynamics}), cycling between the initial and solution states.
However, the dominant avoided level crossing structure is not present in the spin glasses to provide this mechanism.

We now show that there is a very generic mechanism (also described independently by \cite{Hastings19a}) 
that relies on starting in the ground state of the quantum walk part of Hamiltonian $\hat{H}_G$.
Let $\langle \hat{O}\rangle_{\psi(t)}$ for operator $\hat{O}$ be defined by   
$\langle\psi(t)|\hat{O}|\psi(t)\rangle = \langle \hat{O}\rangle_{\psi(t)}$. 
Then, by linearity, and the definition of $\hat{H}(\gamma)$ in (\ref{eq:Hqwcomp}), the energy expectation at time $t$ is 
\begin{eqnarray}
\langle \hat{H}(\gamma)\rangle_{\psi(t)} &=& \langle \hat{H}_G\rangle_{\psi(t)} + \langle \hat{H}_P\rangle_{\psi(t)}.
\end{eqnarray}
Due to the unitarity of the evolution under a time-independent Hamiltonian, this expectation energy will not change over time, giving 
\begin{eqnarray}
\langle \hat{H}(\gamma)\rangle_{\psi(t)} &=& \langle \hat{H}(\gamma)\rangle_{\psi(0)}.
\end{eqnarray}
which yields 
\begin{eqnarray}
\langle \hat{H}_G\rangle_{\psi(t)} - \langle \hat{H}_G\rangle_{\psi(0)} &=& \langle \hat{H}_P\rangle_{\psi(0)} - \langle \hat{H}_P\rangle_{\psi(t)}.
\end{eqnarray}
As $|\psi(0)\rangle$ is chosen to be the ground state of $\hat{H}_G$, the LHS must be non-negative. 
Furthermore, as $|\psi(0)\rangle$ is not an eigenstate of $\hat{H}(\gamma)$, some dynamics are guaranteed to occur and so the LHS must become positive at early times. 
Therefore, the RHS must also be non-negative always and positive at early times. 
Thus, taking any final time $t_f$, we get the inequality
\begin{eqnarray}\label{eq:econv}
\frac{1}{t_f}\intop_{t=0}^{t_f}\mathrm{d}t\langle \hat{H}_P\rangle_{\psi(t)} &<& \langle \hat{H}_P\rangle_{\psi(0)}.
\end{eqnarray}

Equation (\ref{eq:econv}) shows that performing time evolution under the computational quantum walk Hamiltonian from the initial state $\ket{\psi(0)}$ is guaranteed to lower the energy of the system with respect to $\hat{H}_P$ (the expectation value $\langle \hat{H}_P\rangle_{\psi(t)}$). 
This implies that the overlap with low energy eigenstates of $\hat{H}_P$ will increase, at least for short times.  
A measurement in the computational basis will thus be on average more likely than a random guess to produce a low energy state.

Starting in a low energy state is thus important for the success of the quantum
walk algorithm (we have checked this numerically).  
It also implies that encoding prior information into the initial state will help, 
provided this is given in the form of a lower energy state than the uniform superposition 
state.  
This could be the final state from a previous run, for example, which will be explored further in \cite{Nita2020forthcoming}.
It is also necessary to bias the quantum walk Hamiltonian so that its ground state matches this biased initial state.  
Since this starting state is a known computational basis state, it is possible to do this biasing for suitably designed hardware.

For many optimization problem applications, it is helpful to find a low energy state, even if it is not actually the true ground state. 
From this point of view, that quantum walks necessarily lower the expectation energy with respect to the problem Hamiltonian is very appealing as a computational mechanism.
This argument by itself does not provide a guaranteed scaling or quantum speed up,
but it does explain how the quantum walk dynamics work in this setting, where there is no way to lose (or gain) energy.  
It is possible to generalise these arguments beyond time-independent Hamiltonians \citep{Callison2020forthcoming}, to include monotonic functions $A(t)$ and $B(t)$ in (\ref{eq:TIM}).  

\begin{figure}
  \centering
  \subfigure[]{\includegraphics[width=0.51\textwidth]{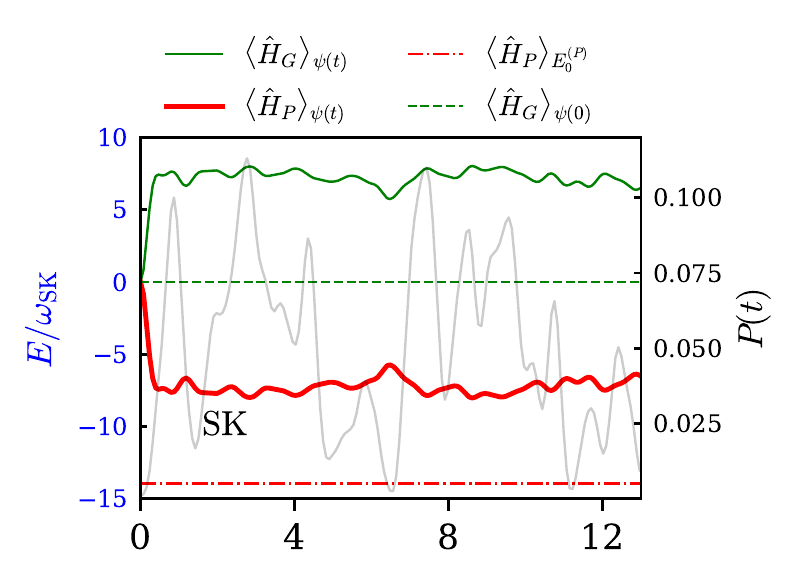}\label{sknrg}}
  \subfigure[]{\includegraphics[width=0.48\textwidth]{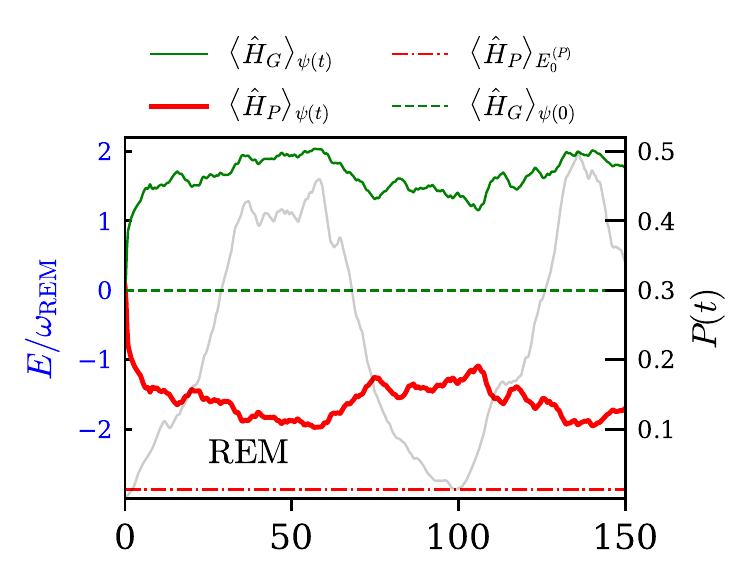}\label{remnrg}}
  \caption{\label{sknrgremnrg}
  The expectation value $\langle \hat{H}_G \rangle_\mathrm{\psi(t)}$ of the quantum walk Hamiltonian (green, thin solid line) and the expectation value $\langle \hat{H}_P \rangle_\mathrm{\psi(t)}$ of the problem Hamiltonian (red, thick solid line) for a typical 10 qubit (a) SK and (b) REM instance.
  The ground state energy eigenvalues of the quantum walk Hamiltonian (green, dashed line) and problem Hamiltonians (red, dash-dotted line) are also shown.
  To illustrate that significant dynamics take place over the timescales used, the instantaneous probabilities $P(t)$ are also shown (grey, faint line).
  The energy values are on the left axes, while probability values are on the right axes.}
\end{figure}

To illustrate this energy redistribution mechanism, the plots in figure \ref{sknrgremnrg} show how the expectation value $\langle\hat{H}_G\rangle_{\psi(t)}$ of the quantum walk Hamiltonian (green solid-line) and the expectation value $\langle\hat{H}_P\rangle_{\psi(t)}$ of the problem Hamiltonian (red solid-line) vary during a quantum walk. 
We have included the instantaneous success probability $P(t)$ (faint grey) to show that the timescale used is long enough for significant dynamics to take place.
A typical 10-qubit SK example is shown in figure \ref{sknrg} and a typical 10-qubit REM example is shown in figure \ref{remnrg}, both on the hypercube using their respective optimal hopping rates $\gamma^{(h)}_\mathrm{opt}$.
Also shown is the ground state eigenvalue $\langle\hat{H}_P\rangle_{E^{(P)}_0}$ of the problem Hamiltonian (red, dash-dotted line) and the ground state eigenvalue $\langle\hat{H}_G\rangle_{\psi(0)}$ of the quantum walk Hamiltonian (green, dashed line).
In both SK and REM, the initial evolution takes the state away from the $\hat{H}_G$ ground state, raising the $\hat{H}_G$ expectation value, and thereby lowering the $\hat{H}_P$ expectation value to a point around which it fluctuates for the duration simulated.  This clearly shows the energy redistribution mechanism at work, and the short time scale over which it appears.
%

%%%%%%%%%%%%%%%%%%%%%%%%%%%%%%%%%%%%%%%%%%%%%%%%%%%%%%%%%%%%%%%%%%%%%%%%%%
\section{Summary and outlook}\label{sec:outlook}
%------------------------------------------------------------------------%

In this work, we have shown numerically that continuous-time quantum walks are a viable computational method for finding ground states of hard spin glass problems. 
We have produced strong numerical evidence for a better-than-search polynomial quantum speed up over random guessing, with a scaling of the average single run success probability ${\sim \tilde{O}(N^{-0.41})}$,
using data sets of size $5\le n\le20$ spins ($32\le N\le 1,048,576$).
Moreover, and importantly, this is obtained without the need to set parameters exponentially precisely, as is required for quantum walk search algorithms. 
The hopping rate $\gamma$, that determines the relative strengths of the quantum walk and problem Hamiltonians, can be estimated from the overall energy scales, which are determined by the hardware and encoding of the problem.   

To explain why quantum walks are able to do better than quantum searching in this case, we compared variants on the spin glass problems that remove or add pairwise correlations, and compared the hypercube graph quantum walk Hamiltonian with the complete graph quantum walk Hamiltonian. 
This showed that the combination of pairwise correlations in the encoding of the problem and a matching single spin flip quantum walk Hamiltonian is required to exploit the correlations.  
The single spin-flips driven by the transverse field terms $\hat{X}_j$ in the hypercube quantum walk Hamiltonian are the correct operators for the pairwise interaction terms $\hat{Z}_j\hat{Z}_k$ in the spin glass Hamiltonian. 
A single spin flip on either qubit $j$ or $k$ changes the energy for that term from high to low, or vice versa.
Since we can choose how to encode the problems into the Hamiltonians, and there are known methods to convert higher order terms to pairwise terms \citep{bremner2002practical,Dattani2019}, we can arrange to use this mechanism both for its computational advantages and practicality for hardware implementation as the transverse Ising Hamiltonian.

To explain how quantum walks are able to find low energy states when the closed quantum dynamics have no mechanism for losing energy, we showed how starting in the ground state of the quantum walk part of the Hamiltonian guarantees dynamics that decrease the expectation value of the energy with respect to the problem Hamiltonian.
This also ensures that prior information can be provided by starting in lower energy states, from which improved solutions can be found.
Exploiting this process will allow an optimal quantum algorithm to be built from multiple quantum walk runs that use the information gained from prior runs.
Performing multiple quantum walk runs in early, noisy quantum hardware is a more viable approach than maintaining coherence for sufficiently accurate adiabatic algorithms. 
Quantum walks may also be simpler to implement since they do not require time dependent controls.  
This work thus provides a significant advance in understanding how to exploit quantum walks in practical hardware for optimization problems.

It is likely that further insights into the computational effectiveness of quantum walks in this transverse Ising Hamiltonian setting are to be found in current knowledge of spin glass phases in the presence of transverse fields.  
The spin glass transition itself is not fully understood, in neither the quantum nor classical case \citep[see, e.g.,][]{Parisi80repSym,Fisher87aSG,Fisher88aSG,Thirumalai89,Larson13aSG,Young17a,Magalhaes17a}.
However, the phases of interest for computation are not the spin glass phases themselves, but the phases where transitions between states are still occurring at a rapid enough rate to find solution states.
Extremely long equilibration timescales are a defining property of all glass phases, including spin glasses \citep{Bouchaud97aGlass,Cugliandolo02Glassy}. 
Since the equilibration (mixing) times $\tau_\mathrm{mix}^{(\epsilon)}$ we find in section \ref{ssec:mixtime} for the SK spin glass only scale polynomially with the number of spins, it is most likely that at the optimal hopping rates $\gamma^{(h)}_\mathrm{opt}$, our quantum walks are not in a finite size precursor to a spin glass phase, but rather in a precursor to a paramagnetic phase, for which equilibration times can be fast. 
Given that the system should localize more in lower energy states for smaller transverse fields, it is reasonable that our optimal hopping rates $\gamma^{(h)}_\mathrm{opt}$ occur near the edge of the precursor to the spin glass phase. 
Furthermore, the mild scaling of the width $\Delta \gamma^{(h)}_\mathrm{opt}$ of the peak around the optimal hopping rate $\gamma^{(h)}_\mathrm{opt}$ suggests that the regime where quantum walks performs well may correspond to the second paramagnetic phase observed in \cite{Magalhaes17a}. 
Polynomial gaps have been found around the spin glass--paramagnetic phase transition in a related model in \citep{knysh2016zero}. 

A numerical study such as this inevitably leaves open questions regarding the asymptotic scaling of the problems.  
In particular, we observed a range of hardness in the SK data sets and future work will investigate what fraction of the instances are actually hard for classical algorithms. 
Forthcoming work applying similar techniques to Max2SAT \citep{callison2020max2sat} will characterise the hardness of small random instances in more detail, and establish quantum walks as an effective tool for hard optimization problems more generally. 
While general methods are known to speed up the best classical algorithms \citep{hartwig1984} for this type of problem \citep{Montanaro2015,Montanaro2019SK}, further work is required to determine whether an optimal continuous-time quantum walk algorithm can be devised that fully leverages the advantage from the correlations.
Nonetheless, our work represents a significant advance in developing continuous-time quantum walk computation for hard optimization problems, and provides key insights into the computational mechanisms that can be exploited over short timescales, well-suited to the limited coherence times of noisy, intermediate scale quantum hardware.

%%%%%%%%%%%%%%%%%%%% provides a paragraph for acknowledgements %%%%%%%%%%%%%
\ack
VK and NC funded by UK EPSRC fellowship EP/L022303/1 and NC funded by EPSRC grant EP/S00114X/1. 
AC funded by EPSRC grant EP/L016524/1 via the Imperial College London CDT in Controlled Quantum Dynamics. 
We thank Prof.~Ifan G. Hughes and Dr Ashley Montanaro for helpful discussions.
%%%

%%%%%%%%%%%%%%%%%%%%%%% or can include the generated .bbl file here for submission
\bibliography{qwspinglass} 

\begin{thebibliography}{89}
\providecommand{\natexlab}[1]{#1}
\providecommand{\url}[1]{\texttt{#1}}
\expandafter\ifx\csname urlstyle\endcsname\relax
  \providecommand{\doi}[1]{doi: #1}\else
  \providecommand{\doi}{doi: \begingroup \urlstyle{rm}\Url}\fi

\bibitem[Aharonov et~al.(2001)Aharonov, Ambainis, Kempe, and
  Vazirani]{aharonov00a}
Dorit Aharonov, Andris Ambainis, Julia Kempe, and Umesh Vazirani.
\newblock Quantum walks on graphs.
\newblock In \emph{Proceedings of the Thirty-third Annual ACM Symposium on
  Theory of Computing}, STOC '01, pages 50--59, New York, NY, USA, 2001. ACM.
\newblock ISBN 1-58113-349-9.
\newblock \doi{10.1145/380752.380758}.
\newblock URL \url{http://doi.acm.org/10.1145/380752.380758}.

\bibitem[Ambainis et~al.(2019)Ambainis, Balodis, Iraids, Kokainis, Pr\=usis,
  and Vihrovs]{ambainis2019}
Andris Ambainis, Kaspars Balodis, J\=anis Iraids, Martins Kokainis,
  Kri\v{s}j\=anis Pr\=usis, and Jevg\=enijs Vihrovs.
\newblock \emph{Quantum Speedups for Exponential-Time Dynamic Programming
  Algorithms}, pages 1783--1793.
\newblock Assoc.~for Comp. Machinery, New York, 2019.
\newblock \doi{10.1137/1.9781611975482.107}.
\newblock URL \url{https://epubs.siam.org/doi/abs/10.1137/1.9781611975482.107}.

\bibitem[Amin et~al.(2018)Amin, Andriyash, Rolfe, Kulchytskyy, and
  Melko]{Amin18a}
Mohammad~H. Amin, Evgeny Andriyash, Jason Rolfe, Bohdan Kulchytskyy, and Roger
  Melko.
\newblock Quantum {B}oltzmann machine.
\newblock \emph{Phys. Rev. X}, 8:\penalty0 021050, May 2018.
\newblock \doi{10.1103/PhysRevX.8.021050}.
\newblock URL \url{https://link.aps.org/doi/10.1103/PhysRevX.8.021050}.

\bibitem[Baldwin and Laumann(2018)]{Baldwin2018}
C.~L. Baldwin and C.~R. Laumann.
\newblock Quantum algorithm for energy matching in hard optimization problems.
\newblock \emph{Phys. Rev. B}, 97:\penalty0 224201, Jun 2018.
\newblock \doi{10.1103/PhysRevB.97.224201}.
\newblock URL \url{https://link.aps.org/doi/10.1103/PhysRevB.97.224201}.

\bibitem[Beier and V\"{o}cking(2004)]{Beier2004a}
Rene Beier and Berthold V\"{o}cking.
\newblock Random knapsack in expected polynomial time.
\newblock \emph{Journal of Computer and System Sciences}, 69\penalty0
  (3):\penalty0 306 -- 329, 2004.
\newblock ISSN 0022-0000.
\newblock \doi{https://doi.org/10.1016/j.jcss.2004.04.004}.
\newblock URL
  \url{http://www.sciencedirect.com/science/article/pii/S0022000004000431}.
\newblock Special Issue on STOC 2003.

\bibitem[Bennett et~al.(1997)Bennett, Bernstein, Brassard, and
  Vazirani]{bennett1997strengths}
C.~Bennett, E.~Bernstein, G.~Brassard, and U.~Vazirani.
\newblock Strengths and {W}eaknesses of {Q}uantum {C}omputing.
\newblock \emph{SIAM Journal on Computing}, 26\penalty0 (5):\penalty0
  1510--1523, 1997.
\newblock \doi{10.1137/S0097539796300933}.
\newblock URL \url{https://doi.org/10.1137/S0097539796300933}.

\bibitem[Bernien et~al.(2017)Bernien, Schwartz, Keesling, Levine, Omran,
  Pichler, Choi, Zibrov, Endres, Greiner, Vuleti{\'c}, and Lukin]{Bernien17a}
Hannes Bernien, Sylvain Schwartz, Alexander Keesling, Harry Levine, Ahmed
  Omran, Hannes Pichler, Soonwon Choi, Alexander~S. Zibrov, Manuel Endres,
  Markus Greiner, Vladan Vuleti{\'c}, and Mikhail~D. Lukin.
\newblock Probing many-body dynamics on a 51-atom quantum simulator.
\newblock \emph{Nature}, 551:\penalty0 579 EP --, 11 2017.
\newblock URL \url{https://doi.org/10.1038/nature24622}.

\bibitem[Bian et~al.(2013)Bian, Chudak, Macready, Clark, and Gaitan]{Bian13a}
Zhengbing Bian, Fabian Chudak, William~G. Macready, Lane Clark, and Frank
  Gaitan.
\newblock Experimental {D}etermination of {R}amsey {N}umbers.
\newblock \emph{Phys. Rev. Lett.}, 111:\penalty0 130505, Sep 2013.
\newblock \doi{10.1103/PhysRevLett.111.130505}.
\newblock URL \url{https://link.aps.org/doi/10.1103/PhysRevLett.111.130505}.

\bibitem[Boixo et~al.(2013)Boixo, Albash, Spedalieri, Chancellor, and
  Lidar]{boixo2013experimental}
Sergio Boixo, Tameem Albash, Federico~M. Spedalieri, Nicholas Chancellor, and
  Daniel~A. Lidar.
\newblock Experimental signature of programmable quantum annealing.
\newblock \emph{Nature Communications}, 4:\penalty0 2067 EP --, 06 2013.
\newblock URL \url{https://doi.org/10.1038/ncomms3067}.

\bibitem[Bouchaud et~al.(1998)Bouchaud, Cugliandolo, Kurchan, and
  Mezard]{Bouchaud97aGlass}
Jean-Philippe Bouchaud, Leticia~F Cugliandolo, Jorge Kurchan, and Marc Mezard.
\newblock \emph{Out of equilibrium dynamics in spin-glasses and other glassy
  systems}, pages 161--223.
\newblock World Scientific, Singapore, 1998.
\newblock \doi{10.1142/9789812819437_0006}.
\newblock URL
  \url{https://www.worldscientific.com/doi/abs/10.1142/9789812819437_0006}.

\bibitem[Bremner et~al.(2002)Bremner, Dawson, Dodd, Gilchrist, Harrow,
  Mortimer, Nielsen, and Osborne]{bremner2002practical}
Michael~J. Bremner, Christopher~M. Dawson, Jennifer~L. Dodd, Alexei Gilchrist,
  Aram~W. Harrow, Duncan Mortimer, Michael~A. Nielsen, and Tobias~J. Osborne.
\newblock Practical {S}cheme for {Q}uantum {C}omputation with {A}ny
  {T}wo-{Q}ubit {E}ntangling {G}ate.
\newblock \emph{Phys. Rev. Lett.}, 89:\penalty0 247902, Nov 2002.
\newblock \doi{10.1103/PhysRevLett.89.247902}.
\newblock URL \url{https://link.aps.org/doi/10.1103/PhysRevLett.89.247902}.

\bibitem[Callison et~al.(2020{\natexlab{a}})Callison, Festenstein, Light,
  Chancellor, and Kendon]{callison2020max2sat}
Adam Callison, Max Festenstein, Lewis Light, Nicholas Chancellor, and Viv
  Kendon.
\newblock Hybrid adiabatic-quantum-walk algorithms applied to {Max2SAT}, in
  preparation., 2020{\natexlab{a}}.
\newblock in preparation.

\bibitem[Callison et~al.(2020{\natexlab{b}})Callison, Kendon, and
  Chancellor]{Callison2020forthcoming}
Adam Callison, Viv Kendon, and Nicholas Chancellor.
\newblock Tools for practical quantum annealing, 2020{\natexlab{b}}.
\newblock in preparation.

\bibitem[Chancellor et~al.(2016)Chancellor, Zohren, Warburton, Benjamin, and
  Roberts]{chancellor16a}
N.~Chancellor, S.~Zohren, P.~A. Warburton, S.~C. Benjamin, and S.~Roberts.
\newblock A direct mapping of {M}ax k-{SAT} and high order parity checks to a
  chimera graph.
\newblock \emph{Scientific Reports}, 6:\penalty0 37107 EP --, 11 2016.
\newblock URL \url{https://doi.org/10.1038/srep37107}.

\bibitem[Chancellor et~al.(2019)Chancellor, Callison, Kendon, and
  Mintert]{data_arch}
N~Chancellor, A~Callison, V~Kendon, and F~Mintert.
\newblock Finding spin-glass ground states using quantum walks [dataset], 2019.
\newblock URL \url{https://doi.org/10.15128/r21544bp097}.
\newblock Data archive at Durham University, UK, for spin glass instances used
  in this work.

\bibitem[Chancellor(2017)]{chancellor2017modernizing}
Nicholas Chancellor.
\newblock Modernizing quantum annealing using local searches.
\newblock \emph{New Journal of Physics}, 19\penalty0 (2):\penalty0 023024, feb
  2017.
\newblock \doi{10.1088/1367-2630/aa59c4}.
\newblock URL \url{https://doi.org/10.1088/1367-2630/aa59c4}.

\bibitem[Chancellor et~al.(2017)Chancellor, Zohren, and
  Warburton]{chancellor17a}
Nicholas Chancellor, Stefan Zohren, and Paul~A. Warburton.
\newblock Circuit design for multi-body interactions in superconducting quantum
  annealing systems with applications to a scalable architecture.
\newblock \emph{npj Quantum Information}, 3\penalty0 (21), 2017.
\newblock \doi{10.1038/s41534-017-0022-6}.
\newblock URL \url{https://www.nature.com/articles/s41534-017-0022-6}.

\bibitem[Childs and Goldstone(2004)]{childs2004spatial}
Andrew~M. Childs and Jeffrey Goldstone.
\newblock Spatial search by quantum walk.
\newblock \emph{Phys. Rev. A}, 70:\penalty0 022314, Aug 2004.
\newblock \doi{10.1103/PhysRevA.70.022314}.
\newblock URL \url{https://link.aps.org/doi/10.1103/PhysRevA.70.022314}.

\bibitem[Childs et~al.(2003)Childs, Cleve, Deotto, Farhi, Gutmann, and
  Spielman]{childs2003exponential}
Andrew~M. Childs, Richard Cleve, Enrico Deotto, Edward Farhi, Sam Gutmann, and
  Daniel~A. Spielman.
\newblock Exponential {A}lgorithmic {S}peedup by a {Q}uantum {W}alk.
\newblock In \emph{Proceedings of the Thirty-fifth Annual ACM Symposium on
  Theory of Computing}, STOC '03, pages 59--68, New York, NY, USA, 2003. ACM.
\newblock ISBN 1-58113-674-9.
\newblock \doi{10.1145/780542.780552}.
\newblock URL \url{http://doi.acm.org/10.1145/780542.780552}.

\bibitem[Choi(2010)]{choi2010adiabatic}
Vicky Choi.
\newblock Adiabatic quantum algorithms for the {NP}-complete {M}aximum-{W}eight
  {I}ndependent set, {E}xact {C}over and 3{SAT} problems, 2010.
\newblock URL \url{https://arxiv.org/abs/1004.2226}.
\newblock arXiv:1004.2226.

\bibitem[{Coxson} et~al.(2014){Coxson}, {Hill}, and {Russo}]{coxson14a}
G.~E. {Coxson}, C.~R. {Hill}, and J.~C. {Russo}.
\newblock Adiabatic quantum computing for finding low-peak-sidelobe codes.
\newblock In \emph{2014 IEEE High Performance Extreme Computing Conference
  (HPEC)}, pages 1--6, Sep. 2014.
\newblock \doi{10.1109/HPEC.2014.7040953}.
\newblock URL \url{https://ieeexplore.ieee.org/abstract/document/7040953}.

\bibitem[Cugliandolo(2002)]{Cugliandolo02Glassy}
Leticia~F Cugliandolo.
\newblock Dynamics of glassy systems, 2002.
\newblock URL \url{https://arxiv.org/abs/cond-mat/0210312}.
\newblock arXiv preprint cond-mat/0210312.

\bibitem[D-{W}ave(1999--)]{D-wave}
D-{W}ave.
\newblock {D}-{W}ave {S}ystems {I}nc. website, 1999--.
\newblock URL \url{http://www.dwavesys.com/}.
\newblock [Online; accessed \today].

\bibitem[Dalzell et~al.(2017)Dalzell, Yoder, and Chuang]{Dalzell2017fixedpoint}
Alexander~M. Dalzell, Theodore~J. Yoder, and Isaac~L. Chuang.
\newblock Fixed-point adiabatic quantum search.
\newblock \emph{Phys. Rev. A}, 95:\penalty0 012311, Jan 2017.
\newblock \doi{10.1103/PhysRevA.95.012311}.
\newblock URL \url{https://link.aps.org/doi/10.1103/PhysRevA.95.012311}.

\bibitem[Dattani(2019)]{Dattani2019}
Nike Dattani.
\newblock Quadratization in discrete optimization and quantum mechanics, 2019.
\newblock URL \url{http://arxiv.org/abs/1901.04405}.
\newblock arXiv:1901.04405.

\bibitem[De~las Cuevas and Cubitt(2016)]{de2016simple}
Gemma De~las Cuevas and Toby~S. Cubitt.
\newblock Simple universal models capture all classical spin physics.
\newblock \emph{Science}, 351\penalty0 (6278):\penalty0 1180--1183, 2016.
\newblock ISSN 0036-8075.
\newblock \doi{10.1126/science.aab3326}.
\newblock URL \url{http://science.sciencemag.org/content/351/6278/1180}.

\bibitem[Derrida(1980)]{derrida1980random}
B.~Derrida.
\newblock Random-{E}nergy {M}odel: {L}imit of a {F}amily of {D}isordered
  {M}odels.
\newblock \emph{Phys. Rev. Lett.}, 45:\penalty0 79--82, Jul 1980.
\newblock \doi{10.1103/PhysRevLett.45.79}.
\newblock URL \url{https://link.aps.org/doi/10.1103/PhysRevLett.45.79}.

\bibitem[Dodds et~al.(2019)Dodds, Kendon, Adams, and Chancellor]{dodds18a}
A.~Ben Dodds, Viv Kendon, Charles~S. Adams, and Nicholas Chancellor.
\newblock Practical designs for permutation-symmetric problem hamiltonians on
  hypercubes, Sep 2019.
\newblock URL \url{https://link.aps.org/doi/10.1103/PhysRevA.100.032320}.

\bibitem[Duan et~al.(2013)Duan, Zhang, Wu, and Chen]{Duan2013AQC}
Qian-Heng Duan, Shuo Zhang, Wei Wu, and Ping-Xing Chen.
\newblock An alternative approach to construct the initial hamiltonian of the
  adiabatic quantum computation.
\newblock \emph{Chinese Physics Letters}, 30\penalty0 (1):\penalty0 010302, Jan
  2013.
\newblock \doi{10.1088/0256-307x/30/1/010302}.
\newblock URL \url{https://doi.org/10.1088/0256-307x/30/1/010302}.

\bibitem[Farhi and Gutmann(1998)]{farhi98a}
Edward Farhi and Sam Gutmann.
\newblock Quantum computation and decision trees.
\newblock \emph{Phys. Rev. A}, 58:\penalty0 915--928, Aug 1998.
\newblock \doi{10.1103/PhysRevA.58.915}.
\newblock URL \url{http://link.aps.org/doi/10.1103/PhysRevA.58.915}.

\bibitem[Farhi et~al.(2000)Farhi, Goldstone, Gutmann, and
  Sipser]{farhi2000quantum}
Edward Farhi, Jeffrey Goldstone, Sam Gutmann, and Michael Sipser.
\newblock Quantum computation by adiabatic evolution, 2000.
\newblock URL \url{https://arxiv.org/abs/quant-ph/0001106}.
\newblock arXiv preprint quant-ph/0001106.

\bibitem[Farhi et~al.(2001)Farhi, Goldstone, Gutmann, Lapan, Lundgren, and
  Preda]{farhi2001quantum}
Edward Farhi, Jeffrey Goldstone, Sam Gutmann, Joshua Lapan, Andrew Lundgren,
  and Daniel Preda.
\newblock A {Q}uantum {A}diabatic {E}volution {A}lgorithm {A}pplied to {R}andom
  {I}nstances of an {NP}-{C}omplete {P}roblem.
\newblock \emph{Science}, 292\penalty0 (5516):\penalty0 472--475, 2001.
\newblock ISSN 0036-8075.
\newblock \doi{10.1126/science.1057726}.
\newblock URL \url{http://science.sciencemag.org/content/292/5516/472}.

\bibitem[Farhi et~al.(2008)Farhi, Goldstone, Gutmann, and Nagaj]{farhi2008make}
Edward Farhi, Jeffrey Goldstone, Sam Gutmann, and Daniel Nagaj.
\newblock How to make the quantum adiabatic algorithm fail.
\newblock \emph{International Journal of Quantum Information}, 06\penalty0
  (03):\penalty0 503--516, 2008.
\newblock \doi{10.1142/S021974990800358X}.
\newblock URL \url{https://doi.org/10.1142/S021974990800358X}.

\bibitem[Farhi et~al.(2011)Farhi, Goldstone, Gosset, Gutmann, and
  Shor]{farhi2011unstructured}
Edward Farhi, Jeffrey Goldstone, David Gosset, Sam Gutmann, and Peter Shor.
\newblock Unstructured randomness, small gaps and localization.
\newblock \emph{Quantum Information \& Computation}, 11\penalty0
  (9-10):\penalty0 840--854, Sep 2011.
\newblock URL
  \url{http://www.rintonpress.com/xxqic11/qic-11-910/0840-0854.pdf}.

\bibitem[Farhi et~al.(2014{\natexlab{a}})Farhi, Goldstone, and
  Gutmann]{Farhi14a}
Edward Farhi, Jeffrey Goldstone, and Sam Gutmann.
\newblock A quantum approximate optimization algorithm, 2014{\natexlab{a}}.
\newblock URL \url{https://arxiv.org/abs/1411.4028}.
\newblock arXiv:1411.4028.

\bibitem[Farhi et~al.(2014{\natexlab{b}})Farhi, Goldstone, and
  Gutmann]{Farhi14b}
Edward Farhi, Jeffrey Goldstone, and Sam Gutmann.
\newblock A quantum approximate optimization algorithm applied to a bounded
  occurrence constraint problem, 2014{\natexlab{b}}.
\newblock URL \url{https://arxiv.org/abs/1412.6062}.
\newblock arXiv:1412.6062.

\bibitem[Feng et~al.(2014)Feng, Fang, Tam, Yun, Ramanujam, Moreno, and
  Jarrell]{Feng14a}
Sheng Feng, Ye~Fang, Ka-Ming Tam, Zhifeng Yun, J~Ramanujam, Juana Moreno, and
  Mark Jarrell.
\newblock Three {D}imensional {E}dwards-{A}nderson {S}pin {G}lass {M}odel in an
  {E}xternal {F}ield, 2014.
\newblock URL \url{https://arxiv.org/abs/1403.4560}.
\newblock arXiv:1403.4560.

\bibitem[Finnila et~al.(1994)Finnila, Gomez, Sebenik, Stenson, and
  Doll]{Fini1994}
A.B. Finnila, M.A. Gomez, C.~Sebenik, C.~Stenson, and J.D. Doll.
\newblock Quantum annealing: {A} new method for minimizing multidimensional
  functions.
\newblock \emph{Chemical Physics Letters}, 219\penalty0 (5):\penalty0 343 --
  348, 1994.
\newblock ISSN 0009-2614.
\newblock \doi{https://doi.org/10.1016/0009-2614(94)00117-0}.
\newblock URL
  \url{http://www.sciencedirect.com/science/article/pii/0009261494001170}.

\bibitem[Fisher and Huse(1987)]{Fisher87aSG}
D~S Fisher and D~A Huse.
\newblock Absence of many states in realistic spin glasses.
\newblock \emph{Journal of Physics A: Mathematical and General}, 20\penalty0
  (15):\penalty0 L1005--L1010, oct 1987.
\newblock \doi{10.1088/0305-4470/20/15/013}.
\newblock URL \url{https://doi.org/10.1088/0305-4470/20/15/013}.

\bibitem[Fisher and Huse(1988)]{Fisher88aSG}
Daniel~S. Fisher and David~A. Huse.
\newblock Equilibrium behavior of the spin-glass ordered phase.
\newblock \emph{Phys. Rev. B}, 38:\penalty0 386--411, Jul 1988.
\newblock \doi{10.1103/PhysRevB.38.386}.
\newblock URL \url{https://link.aps.org/doi/10.1103/PhysRevB.38.386}.

\bibitem[Gra\ss{}(2019)]{Grass2019longitudinal}
Tobias Gra\ss{}.
\newblock Quantum annealing with longitudinal bias fields.
\newblock \emph{Phys. Rev. Lett.}, 123:\penalty0 120501, Sep 2019.
\newblock \doi{10.1103/PhysRevLett.123.120501}.
\newblock URL \url{https://link.aps.org/doi/10.1103/PhysRevLett.123.120501}.

\bibitem[Gra\ss{} and Lewenstein(2017)]{Grass2017hybrid}
Tobias Gra\ss{} and Maciej Lewenstein.
\newblock Hybrid annealing: Coupling a quantum simulator to a classical
  computer.
\newblock \emph{Phys. Rev. A}, 95:\penalty0 052309, May 2017.
\newblock \doi{10.1103/PhysRevA.95.052309}.
\newblock URL \url{https://link.aps.org/doi/10.1103/PhysRevA.95.052309}.

\bibitem[Grover(1996)]{grover1996fast}
Lov~K. Grover.
\newblock A fast quantum mechanical algorithm for database search.
\newblock In \emph{Proceedings of the Twenty-eighth Annual ACM Symposium on
  Theory of Computing}, STOC '96, pages 212--219, New York, NY, USA, 1996. ACM.
\newblock ISBN 0-89791-785-5.
\newblock \doi{10.1145/237814.237866}.
\newblock URL \url{http://doi.acm.org/10.1145/237814.237866}.

\bibitem[Hadfield et~al.(2019)Hadfield, Wang, O'Gorman, Rieffel, Venturelli,
  and Biswas]{Hadfield17a}
Stuart Hadfield, Zhihui Wang, Bryan O'Gorman, Eleanor~G. Rieffel, Davide
  Venturelli, and Rupak Biswas.
\newblock From the quantum approximate optimization algorithm to a quantum
  alternating operator ansatz.
\newblock \emph{Algorithms}, 12\penalty0 (2), 2019.
\newblock ISSN 1999-4893.
\newblock \doi{10.3390/a12020034}.
\newblock URL \url{http://www.mdpi.com/1999-4893/12/2/34}.

\bibitem[Hamze et~al.(2019)Hamze, Raymond, Pattison, Biswas, and
  Katzgraber]{Hamze19tunable}
Firas Hamze, Jack Raymond, Christopher~A. Pattison, Katja Biswas, and Helmut~G.
  Katzgraber.
\newblock The wishart planted ensemble: A tunably-rugged pairwise ising model
  with a first-order phase transition, 2019.
\newblock URL \url{https://arxiv.org/abs/1906.00275}.
\newblock arXiv preprint arXiv:1906.00275.

\bibitem[Hartwig et~al.(1984)Hartwig, Daske, and Kobe]{hartwig1984}
A.~Hartwig, F.~Daske, and S.~Kobe.
\newblock A recursive branch-and-bound algorithm for the exact ground state of
  ising spin-glass models.
\newblock \emph{Computer Physics Communications}, 32\penalty0 (2):\penalty0 133
  -- 138, 1984.
\newblock ISSN 0010-4655.
\newblock \doi{https://doi.org/10.1016/0010-4655(84)90066-3}.
\newblock URL
  \url{http://www.sciencedirect.com/science/article/pii/0010465584900663}.

\bibitem[Hastings(2019)]{Hastings19a}
Matthew~B. Hastings.
\newblock Duality in {Q}uantum {Q}uenches and {C}lassical {A}pproximation
  {A}lgorithms: {P}retty {G}ood or {V}ery {B}ad, November 2019.
\newblock ISSN 2521-327X.
\newblock URL \url{https://doi.org/10.22331/q-2019-11-11-201}.

\bibitem[Hen(2014)]{Hen2014Continuous}
Itay Hen.
\newblock Continuous-time quantum algorithms for unstructured problems.
\newblock \emph{Journal of Physics A: Mathematical and Theoretical},
  47\penalty0 (4):\penalty0 045305, Jan 2014.
\newblock \doi{10.1088/1751-8113/47/4/045305}.
\newblock URL \url{https://doi.org/10.1088/1751-8113/47/4/045305}.

\bibitem[Hen(2019)]{Hen19planted}
Itay Hen.
\newblock Equation planting: A tool for benchmarking ising machines.
\newblock \emph{Phys. Rev. Applied}, 12:\penalty0 011003, Jul 2019.
\newblock \doi{10.1103/PhysRevApplied.12.011003}.
\newblock URL \url{https://link.aps.org/doi/10.1103/PhysRevApplied.12.011003}.

\bibitem[Hunter(2007)]{hunter2007matplotlib}
John~D Hunter.
\newblock Matplotlib: A 2{D} graphics environment.
\newblock \emph{Computing in science \& engineering}, 9\penalty0 (3):\penalty0
  90, 2007.

\bibitem[Inagaki et~al.(2016)Inagaki, Haribara, Igarashi, Sonobe, Tamate,
  Honjo, Marandi, McMahon, Umeki, Enbutsu, Tadanaga, Takenouchi, Aihara,
  Kawarabayashi, Inoue, Utsunomiya, and Takesue]{Inagaki16a}
Takahiro Inagaki, Yoshitaka Haribara, Koji Igarashi, Tomohiro Sonobe, Shuhei
  Tamate, Toshimori Honjo, Alireza Marandi, Peter~L. McMahon, Takeshi Umeki,
  Koji Enbutsu, Osamu Tadanaga, Hirokazu Takenouchi, Kazuyuki Aihara, Ken-ichi
  Kawarabayashi, Kyo Inoue, Shoko Utsunomiya, and Hiroki Takesue.
\newblock A coherent {I}sing machine for 2000-node optimization problems.
\newblock \emph{Science}, 354\penalty0 (6312):\penalty0 603--606, 2016.
\newblock ISSN 0036-8075.
\newblock URL \url{http://science.sciencemag.org/content/354/6312/603}.

\bibitem[Johnson et~al.(2011)Johnson, Amin, Gildert, Lanting, Hamze, Dickson,
  Harris, Berkley, Johansson, Bunyk, Chapple, Enderud, Hilton, Karimi,
  Ladizinsky, Ladizinsky, Oh, Perminov, Rich, Thom, Tolkacheva, Truncik,
  Uchaikin, Wang, Wilson, and Rose]{johnson2011quantum}
M.~W. Johnson, M.~H.~S. Amin, S.~Gildert, T.~Lanting, F.~Hamze, N.~Dickson,
  R.~Harris, A.~J. Berkley, J.~Johansson, P.~Bunyk, E.~M. Chapple, C.~Enderud,
  J.~P. Hilton, K.~Karimi, E.~Ladizinsky, N.~Ladizinsky, T.~Oh, I.~Perminov,
  C.~Rich, M.~C. Thom, E.~Tolkacheva, C.~J.~S. Truncik, S.~Uchaikin, J.~Wang,
  B.~Wilson, and G.~Rose.
\newblock Quantum annealing with manufactured spins.
\newblock \emph{Nature}, 473:\penalty0 194 EP --, 05 2011.
\newblock URL \url{https://doi.org/10.1038/nature10012}.

\bibitem[Jones et~al.(2001--)Jones, Oliphant, Peterson, et~al.]{scipy}
Eric Jones, Travis Oliphant, Pearu Peterson, et~al.
\newblock {SciPy}: {O}pen source scientific tools for {Python}, 2001--.
\newblock URL \url{http://www.scipy.org/}.
\newblock [Online; accessed \today].

\bibitem[Jordan and Farhi(2008)]{Jordan08gadget}
Stephen Jordan and Edward Farhi.
\newblock Perturbative gadgets at arbitrary orders.
\newblock \emph{Physical Review A}, 77, 02 2008.
\newblock URL \url{https://link.aps.org/doi/10.1103/PhysRevA.77.062329}.

\bibitem[Kadowaki and Nishimori(1998)]{Kado1998}
Tadashi Kadowaki and Hidetoshi Nishimori.
\newblock Quantum annealing in the transverse {I}sing model.
\newblock \emph{Phys. Rev. E}, 58:\penalty0 5355--5363, Nov 1998.
\newblock \doi{10.1103/PhysRevE.58.5355}.
\newblock URL \url{https://link.aps.org/doi/10.1103/PhysRevE.58.5355}.

\bibitem[Katzgraber et~al.(2014)Katzgraber, Hamze, and
  Andrist]{katzgraber2014glassy}
Helmut~G. Katzgraber, Firas Hamze, and Ruben~S. Andrist.
\newblock Glassy chimeras could be blind to quantum speedup: Designing better
  benchmarks for quantum annealing machines.
\newblock \emph{Phys. Rev. X}, 4:\penalty0 021008, Apr 2014.
\newblock \doi{10.1103/PhysRevX.4.021008}.
\newblock URL \url{https://link.aps.org/doi/10.1103/PhysRevX.4.021008}.

\bibitem[Kechedzhi et~al.(2018)Kechedzhi, Smelyanskiy, McClean, Denchev,
  Mohseni, Isakov, Boixo, Altshuler, and Neven]{kechedzhi18a}
Kostyantyn Kechedzhi, Vadim Smelyanskiy, Jarrod~R. McClean, Vasil~S. Denchev,
  Masoud Mohseni, Sergei Isakov, Sergio Boixo, Boris Altshuler, and Hartmut
  Neven.
\newblock {Efficient Population Transfer via Non-Ergodic Extended States in
  Quantum Spin Glass}.
\newblock In Stacey Jeffery, editor, \emph{13th Conference on the Theory of
  Quantum Computation, Communication and Cryptography (TQC 2018)}, volume 111
  of \emph{Leibniz International Proceedings in Informatics (LIPIcs)}, pages
  9:1--9:16, Dagstuhl, Germany, 2018. Schloss Dagstuhl--Leibniz-Zentrum fuer
  Informatik.
\newblock ISBN 978-3-95977-080-4.
\newblock \doi{10.4230/LIPIcs.TQC.2018.9}.
\newblock URL \url{http://drops.dagstuhl.de/opus/volltexte/2018/9256}.

\bibitem[Kim et~al.(2011)Kim, Korenblit, Islam, Edwards, Chang, Noh,
  Carmichael, Lin, Duan, Wang, Freericks, and Monroe]{Kim2011ionIsing}
K~Kim, S~Korenblit, R~Islam, E~E Edwards, M-S Chang, C~Noh, H~Carmichael, G-D
  Lin, L-M Duan, C~C~Joseph Wang, J~K Freericks, and C~Monroe.
\newblock Quantum simulation of the transverse ising model with trapped ions.
\newblock \emph{New Journal of Physics}, 13\penalty0 (10):\penalty0 105003, oct
  2011.
\newblock \doi{10.1088/1367-2630/13/10/105003}.
\newblock URL \url{https://doi.org/10.1088/1367-2630/13/10/105003}.

\bibitem[Kluyver et~al.(2016)Kluyver, Ragan-Kelley, P{\'e}rez, Granger,
  Bussonnier, Frederic, Kelley, Hamrick, Grout, Corlay, Ivanov, Avila, Abdalla,
  and Willing]{jupyter}
Thomas Kluyver, Benjamin Ragan-Kelley, Fernando P{\'e}rez, Brian Granger,
  Matthias Bussonnier, Jonathan Frederic, Kyle Kelley, Jessica Hamrick, Jason
  Grout, Sylvain Corlay, Paul Ivanov, Dami{\'a}n Avila, Safia Abdalla, and
  Carol Willing.
\newblock Jupyter {N}otebooks -- a publishing format for reproducible
  computational workflows.
\newblock In F.~Loizides and B.~Schmidt, editors, \emph{Positioning and {P}ower
  in {A}cademic {P}ublishing: {P}layers, {A}gents and {A}gendas}, pages 87 --
  90. IOS Press, 2016.

\bibitem[Knysh(2016)]{knysh2016zero}
Sergey Knysh.
\newblock Zero-temperature quantum annealing bottlenecks in the spin-glass
  phase.
\newblock \emph{Nature Communications}, 7:\penalty0 12370 EP --, 08 2016.
\newblock URL \url{https://doi.org/10.1038/ncomms12370}.

\bibitem[Krivelevich and Vilenchik(2006)]{Krivelevich2006a}
Michael Krivelevich and Dan Vilenchik.
\newblock Solving random satisfiable 3{CNF} formulas in expected polynomial
  time.
\newblock In \emph{Proceedings of the Seventeenth Annual ACM-SIAM Symposium on
  Discrete Algorithm}, SODA '06, pages 454--463, Philadelphia, PA, USA, 2006.
  Society for Industrial and Applied Mathematics.
\newblock ISBN 0-89871-605-5.
\newblock URL \url{http://dl.acm.org/citation.cfm?id=1109557.1109608}.

\bibitem[Larson et~al.(2013)Larson, Katzgraber, Moore, and Young]{Larson13aSG}
Derek Larson, Helmut~G. Katzgraber, M.~A. Moore, and A.~P. Young.
\newblock Spin glasses in a field: Three and four dimensions as seen from one
  space dimension.
\newblock \emph{Phys. Rev. B}, 87:\penalty0 024414, Jan 2013.
\newblock \doi{10.1103/PhysRevB.87.024414}.
\newblock URL \url{https://link.aps.org/doi/10.1103/PhysRevB.87.024414}.

\bibitem[Li et~al.(2017)Li, Dattani, Chen, Liu, Wang, Tanburn, Chen, Peng, and
  Du]{Li17a}
Zhaokai Li, Nikesh~S Dattani, Xi~Chen, Xiaomei Liu, Hengyan Wang, Richard
  Tanburn, Hongwei Chen, Xinhua Peng, and Jiangfeng Du.
\newblock High-fidelity adiabatic quantum computation using the intrinsic
  {H}amiltonian of a spin system: Application to the experimental factorization
  of 291311, 2017.
\newblock URL \url{https://arxiv.org/abs/1706.08061}.
\newblock arXiv preprint arXiv:1706.08061.

\bibitem[Lucas(2014)]{lucas2014ising}
Andrew Lucas.
\newblock Ising formulations of many {NP} problems.
\newblock \emph{Frontiers in Physics}, 2:\penalty0 5, 2014.
\newblock ISSN 2296-424X.
\newblock \doi{10.3389/fphy.2014.00005}.
\newblock URL
  \url{https://www.frontiersin.org/article/10.3389/fphy.2014.00005}.

\bibitem[Magalhaes et~al.(2017)Magalhaes, Morais, Zimmer, Lazo, and
  Nobre]{Magalhaes17a}
S.~G. Magalhaes, C.~V. Morais, F.~M. Zimmer, M.~J. Lazo, and F.~D. Nobre.
\newblock Nonlinear susceptibility of a quantum spin glass under uniform
  transverse and random longitudinal magnetic fields.
\newblock \emph{Phys. Rev. B}, 95:\penalty0 064201, Feb 2017.
\newblock \doi{10.1103/PhysRevB.95.064201}.
\newblock URL \url{https://link.aps.org/doi/10.1103/PhysRevB.95.064201}.

\bibitem[Marsh and Wang(2019)]{Marsh2018qwqaoa}
S.~Marsh and J.~B. Wang.
\newblock A quantum walk-assisted approximate algorithm for bounded {NP}
  optimisation problems.
\newblock \emph{Quantum Information Processing}, 18\penalty0 (3):\penalty0 61,
  Jan 2019.
\newblock ISSN 1573-1332.
\newblock \doi{10.1007/s11128-019-2171-3}.
\newblock URL \url{https://doi.org/10.1007/s11128-019-2171-3}.

\bibitem[Marshall et~al.(2019)Marshall, Venturelli, Hen, and
  Rieffel]{Marshall19a}
Jeffrey Marshall, Davide Venturelli, Itay Hen, and Eleanor~G. Rieffel.
\newblock Power of pausing: Advancing understanding of thermalization in
  experimental quantum annealers.
\newblock \emph{Phys. Rev. Applied}, 11:\penalty0 044083, Apr 2019.
\newblock \doi{10.1103/PhysRevApplied.11.044083}.
\newblock URL \url{https://link.aps.org/doi/10.1103/PhysRevApplied.11.044083}.

\bibitem[Marzec(2016)]{marzec16a}
Michael Marzec.
\newblock \emph{Portfolio {O}ptimization: {A}pplications in {Q}uantum
  {C}omputing}, chapter~4, pages 73--106.
\newblock John Wiley \& Sons, Ltd, 2016.
\newblock ISBN 9781118593486.
\newblock \doi{10.1002/9781118593486.ch4}.
\newblock URL
  \url{https://onlinelibrary.wiley.com/doi/abs/10.1002/9781118593486.ch4}.

\bibitem[McKinney(2010)]{mckinney2010data}
Wes McKinney.
\newblock Data structures for statistical computing in python.
\newblock In \emph{Proceedings of the 9th Python in Science Conference}, volume
  445, pages 51--56. Austin, TX, 2010.

\bibitem[McMahon et~al.(2016)McMahon, Marandi, Haribara, Hamerly, Langrock,
  Tamate, Inagaki, Takesue, Utsunomiya, Aihara, Byer, Fejer, Mabuchi, and
  Yamamoto]{McMahon16a}
Peter~L. McMahon, Alireza Marandi, Yoshitaka Haribara, Ryan Hamerly, Carsten
  Langrock, Shuhei Tamate, Takahiro Inagaki, Hiroki Takesue, Shoko Utsunomiya,
  Kazuyuki Aihara, Robert~L. Byer, M.~M. Fejer, Hideo Mabuchi, and Yoshihisa
  Yamamoto.
\newblock A fully programmable 100-spin coherent {I}sing machine with
  all-to-all connections.
\newblock \emph{Science}, 354\penalty0 (6312):\penalty0 614--617, 2016.
\newblock ISSN 0036-8075.
\newblock \doi{10.1126/science.aah5178}.
\newblock URL \url{http://science.sciencemag.org/content/354/6312/614}.

\bibitem[Montanaro(2018)]{Montanaro2015}
Ashley Montanaro.
\newblock Quantum-walk speedup of backtracking algorithms.
\newblock \emph{Theory of Computing}, 14\penalty0 (15):\penalty0 1--24, 2018.
\newblock \doi{10.4086/toc.2018.v014a015}.
\newblock URL \url{http://www.theoryofcomputing.org/articles/v014a015}.
\newblock arXiv:1509.02374.

\bibitem[Montanaro(2019)]{Montanaro2019SK}
Ashley Montanaro.
\newblock Quantum speedup of branch-and-bound algorithms, 2019.
\newblock arXiv:1906.10375.

\bibitem[Morley et~al.(2019)Morley, Chancellor, Bose, and
  Kendon]{morley2017quantum}
James~G. Morley, Nicholas Chancellor, Sougato Bose, and Viv Kendon.
\newblock Quantum search with hybrid adiabatic--quantum-walk algorithms and
  realistic noise.
\newblock \emph{Phys. Rev. A}, 99:\penalty0 022339, Feb 2019.
\newblock \doi{10.1103/PhysRevA.99.022339}.
\newblock URL \url{https://link.aps.org/doi/10.1103/PhysRevA.99.022339}.

\bibitem[Moylett et~al.(2017)Moylett, Linden, and Montanaro]{Moylett2017tsp}
Dominic~J. Moylett, Noah Linden, and Ashley Montanaro.
\newblock Quantum speedup of the traveling-salesman problem for bounded-degree
  graphs.
\newblock \emph{Phys. Rev. A}, 95:\penalty0 032323, Mar 2017.
\newblock \doi{10.1103/PhysRevA.95.032323}.
\newblock URL \url{https://link.aps.org/doi/10.1103/PhysRevA.95.032323}.

\bibitem[Nita et~al.(2020)Nita, Walsh, Chen, Callison, Kendon, and
  Chancellor]{Nita2020forthcoming}
Laur Nita, Matthew Walsh, Jie Chen, Adam Callison, Viv Kendon, and Nicholas
  Chancellor.
\newblock Effectiveness of a general continuous time subroutine for hybrid
  quantum/classical optimisation, 2020.
\newblock in preparation.

\bibitem[Oliphant(2006)]{oliphant2006guide}
Travis~E Oliphant.
\newblock \emph{A guide to {NumPy}}, volume~1.
\newblock Trelgol Publishing USA, 2006.

\bibitem[Parisi(1980)]{Parisi80repSym}
G~Parisi.
\newblock The order parameter for spin glasses: a function on the interval 0-1.
\newblock \emph{Journal of Physics A: Mathematical and General}, 13\penalty0
  (3):\penalty0 1101--1112, mar 1980.
\newblock \doi{10.1088/0305-4470/13/3/042}.
\newblock URL \url{https://doi.org/10.1088/0305-4470/13/3/042}.

\bibitem[Passarelli et~al.(2019)Passarelli, Cataudella, and
  Lucignano]{passarelli19a}
G.~Passarelli, V.~Cataudella, and P.~Lucignano.
\newblock Improving quantum annealing of the ferromagnetic $p$-spin model
  through pausing.
\newblock \emph{Phys. Rev. B}, 100:\penalty0 024302, Jul 2019.
\newblock \doi{10.1103/PhysRevB.100.024302}.
\newblock URL \url{https://link.aps.org/doi/10.1103/PhysRevB.100.024302}.

\bibitem[Perdomo-Ortiz et~al.(2011)Perdomo-Ortiz, Venegas-Andraca, and
  Aspuru-Guzik]{Perdomo-Ortiz11guessing}
Alejandro Perdomo-Ortiz, Salvador~E. Venegas-Andraca, and Al{\'a}n
  Aspuru-Guzik.
\newblock A study of heuristic guesses for adiabatic quantum computation.
\newblock \emph{Quantum Information Processing}, 10\penalty0 (1):\penalty0
  33--52, Feb 2011.
\newblock ISSN 1573-1332.
\newblock \doi{10.1007/s11128-010-0168-z}.
\newblock URL \url{https://doi.org/10.1007/s11128-010-0168-z}.

\bibitem[Perdomo-Ortiz et~al.(2012)Perdomo-Ortiz, Dickson, Drew-Brook, Rose,
  and Aspuru-Guzik]{perdomo-ortiz12a}
Alejandro Perdomo-Ortiz, Neil Dickson, Marshall Drew-Brook, Geordie Rose, and
  Al{\'a}n Aspuru-Guzik.
\newblock Finding low-energy conformations of lattice protein models by quantum
  annealing.
\newblock \emph{Scientific Reports}, 2:\penalty0 571 EP --, 08 2012.
\newblock URL \url{https://doi.org/10.1038/srep00571}.

\bibitem[{Perez} and {Granger}(2007)]{perez2007ipython}
F.~{Perez} and B.~E. {Granger}.
\newblock {IPython}: {A} {S}ystem for {I}nteractive {S}cientific {C}omputing.
\newblock \emph{Computing in Science Engineering}, 9\penalty0 (3):\penalty0
  21--29, May 2007.
\newblock ISSN 1521-9615.
\newblock \doi{10.1109/MCSE.2007.53}.

\bibitem[Shenvi et~al.(2003)Shenvi, Kempe, and Whaley]{shenvi2003quantum}
Neil Shenvi, Julia Kempe, and K.~Birgitta Whaley.
\newblock Quantum random-walk search algorithm.
\newblock \emph{Phys. Rev. A}, 67:\penalty0 052307, May 2003.
\newblock \doi{10.1103/PhysRevA.67.052307}.
\newblock URL \url{https://link.aps.org/doi/10.1103/PhysRevA.67.052307}.

\bibitem[Sherrington and Kirkpatrick(1975)]{kirkpatrick1975solvable}
David Sherrington and Scott Kirkpatrick.
\newblock Solvable model of a spin-glass.
\newblock \emph{Phys. Rev. Lett.}, 35:\penalty0 1792--1796, Dec 1975.
\newblock \doi{10.1103/PhysRevLett.35.1792}.
\newblock URL \url{https://link.aps.org/doi/10.1103/PhysRevLett.35.1792}.

\bibitem[Thirumalai et~al.(1989)Thirumalai, Li, and Kirkpatrick]{Thirumalai89}
D~Thirumalai, Qiang Li, and T~R Kirkpatrick.
\newblock Infinite-range {I}sing spin glass in a transverse field.
\newblock \emph{Journal of Physics A: Mathematical and General}, 22\penalty0
  (16):\penalty0 3339--3349, aug 1989.
\newblock \doi{10.1088/0305-4470/22/16/023}.
\newblock URL \url{https://doi.org/10.1088/0305-4470/22/16/023}.

\bibitem[Van~Rossum and Drake(2003)]{van2003python}
Guido Van~Rossum and Fred~L Drake.
\newblock \emph{Python language reference manual}.
\newblock Network Theory United Kingdom, 2003.

\bibitem[Yoder et~al.(2014)Yoder, Low, and Chuang]{Yoder2014fixedpoint}
Theodore~J. Yoder, Guang~Hao Low, and Isaac~L. Chuang.
\newblock Fixed-point quantum search with an optimal number of queries.
\newblock \emph{Phys. Rev. Lett.}, 113:\penalty0 210501, Nov 2014.
\newblock \doi{10.1103/PhysRevLett.113.210501}.
\newblock URL \url{https://link.aps.org/doi/10.1103/PhysRevLett.113.210501}.

\bibitem[Young(2017)]{Young17a}
A.~P. Young.
\newblock Stability of the quantum {S}herrington-{K}irkpatrick spin glass
  model.
\newblock \emph{Phys. Rev. E}, 96:\penalty0 032112, Sep 2017.
\newblock \doi{10.1103/PhysRevE.96.032112}.
\newblock URL \url{https://link.aps.org/doi/10.1103/PhysRevE.96.032112}.

\bibitem[Young and Katzgraber(2004)]{Young04a}
A.~P. Young and Helmut~G. Katzgraber.
\newblock Absence of an {A}lmeida-{T}houless line in three-dimensional spin
  glasses.
\newblock \emph{Phys. Rev. Lett.}, 93:\penalty0 207203, Nov 2004.
\newblock \doi{10.1103/PhysRevLett.93.207203}.
\newblock URL \url{https://link.aps.org/doi/10.1103/PhysRevLett.93.207203}.

\bibitem[Zhou et~al.(2018)Zhou, Wang, Choi, Pichler, and Lukin]{zhou2018qaoa}
Leo Zhou, Sheng-Tao Wang, Soonwon Choi, Hannes Pichler, and Mikhail~D. Lukin.
\newblock Quantum approximate optimization algorithm: Performance, mechanism,
  and implementation on near-term devices, 2018.
\newblock URL \url{https://arxiv.org/abs/1812.01041}.
\newblock arXiv preprint arXiv:1812.01041.

\end{thebibliography}

%%%%%%%%%%%%%%%%%%%%%%%%%%%%%%%%%%%%%%%%%%%%%%%%%%%%%%%%%%%%%%%%%%%%%%%%%%%%%%%
\end{document}